\documentclass[twocolumn,traditabstract]{aa}
\usepackage{amssymb}
\usepackage{mathptmx}
\usepackage{natbib}
\bibpunct{(}{)}{;}{a}{}{,}
\usepackage[]{graphicx}
\usepackage{lscape}
\usepackage{txfonts}
\usepackage{booktabs}
\usepackage{multirow}
\usepackage{verbatim}
\usepackage{url}
\defcitealias{canameras20}{C20}
\defcitealias{canameras21}{C21}

\begin{document}

\title{HOLISMOKES - XI. Evaluation of supervised neural networks for strong-lens searches in ground-based imaging surveys}
%  \thanks{Table 1 is only available in electronic form at the CDS via anonymous ftp to cdsarc.u-strasbg.fr (130.79.128.5)
%or via {\tt http://cdsweb.u-strasbg.fr/cgi-bin/qcat?J/A+A/}.}

\author{R. Ca\~nameras\inst{1,2}, S. Schuldt\inst{3}, Y. Shu\inst{4}, S. H. Suyu\inst{1,2,5}, S. Taubenberger\inst{1,2}, I. T. Andika\inst{2,1},
  S. Bag\inst{2,1}, K. T. Inoue\inst{6}, A. T. Jaelani\inst{7,8}, L. Leal-Taix\'e\inst{9}, T. Meinhardt\inst{9}, A. Melo\inst{1,2}, A. More\inst{10,11}}

\institute{
Max-Planck-Institut f\"ur Astrophysik, Karl-Schwarzschild-Str. 1, 85748 Garching, Germany \\
{\tt e-mail: rcanameras@mpa-garching.mpg.de}
\and
Technical University of Munich, TUM School of Natural Sciences, Department of Physics, James-Franck-Stra\ss{}e~1, 85748 Garching, Germany
\and
Dipartimento di Fisica, Universit\`a  degli Studi di Milano, via Celoria 16, I-20133 Milano, Italy
\and
Purple Mountain Observatory, No. 10 Yuanhua Road, Nanjing, Jiangsu, 210033, People's Republic of China
\and
Institute of Astronomy and Astrophysics, Academia Sinica, 11F of ASMAB, No.1, Section 4, Roosevelt Road, Taipei 10617, Taiwan
\and
Kindai University, Faculty of Science and Engineering
\and
Astronomy Research Group and Bosscha Observatory, FMIPA, Institut Teknologi Bandung, Jl. Ganesha 10, Bandung 40132, Indonesia
\and
U-CoE AI-VLB, Institut Teknologi Bandung, Jl. Ganesha 10, Bandung 40132, Indonesia 
\and
Technical University of Munich, Department of Informatics, Boltzmann-Str. 3, 85748 Garching, Germany
\and
Kavli Institute for the Physics and Mathematics of the Universe (WPI), UTIAS, The University of Tokyo, Kashiwa, Chiba 277-8583, Japan
\and
The Inter-University Centre for Astronomy and Astrophysics (IUCAA), Post Bag 4, Ganeshkhind, Pune 411007, India
}

\titlerunning{Testing ML-assisted lens finders}

\authorrunning{R. Ca\~nameras et al.} \date{Received / Accepted}

\abstract{While supervised neural networks have become state of the art for identifying the rare strong gravitational lenses from large imaging
  data sets, their selection remains significantly affected by the large number and diversity of nonlens contaminants. This work evaluates and
  compares systematically the performance of neural networks in order to move towards a rapid selection of galaxy-scale strong lenses with minimal
  human input in the era of deep, wide-scale surveys. We used multiband images from PDR2 of the Hyper-Suprime Cam (HSC) Wide survey to build test
  sets mimicking an actual classification experiment, with 189 strong lenses previously found over the HSC footprint and 70,910 nonlens galaxies in
  COSMOS covering  representative lens-like morphologies. Multiple networks were trained on different sets of realistic strong-lens simulations and
  nonlens galaxies, with various architectures and data pre-processing, mainly using the deepest $gri$ bands. Most networks reached excellent area
  under the Receiver Operating Characteristic (ROC) curves on the test set of 71099 objects, and we determined the ingredients to optimize the true
  positive rate for a total number of false positives equal to zero or 10 (TPR$_{\rm 0}$ and TPR$_{\rm 10}$). The overall performances strongly depend
  on the construction of the ground-truth training data and they typically, but not systematically, improve using our baseline residual network
  architecture. TPR$_{\rm 0}$ tends to be higher for ResNets ($\simeq$ 10--40\%) compared to AlexNet-like networks or G-CNNs. Improvements are found
  when applying random shifts to the image centroids and square root stretches to the pixel values, adding $z$ band, or using random viewpoints of
  the original images, but not when adding $g - \alpha i$ difference images (where $\alpha$ is a tuned constant) to subtract emission from the central
  galaxy. The most significant gain is obtained with committees of networks trained on different data sets, and showing a moderate overlap between
  populations of false positives. Nearly-perfect invariance to image quality can be achieved by using realistic PSF models in our lens simulation
  pipeline, and by training networks either with large number of bands, or jointly with the PSF and science frames. Overall, we show the
  possibility to reach a TPR$_{\rm 0}$ as high as 60\% for the test sets under consideration, which opens promising perspectives for pure selection
  of strong lenses without human input using the {\it Rubin} Observatory and other forthcoming ground-based surveys.}


\maketitle

\section{Introduction}
\label{sec:intro}

Galaxy-scale strong gravitational lenses have a number of important roles in characterizing astrophysical processes underlying galaxy mass
assembly and in constraining the cosmological framework in which these galaxies evolve \citep[e.g.,][and references therein]{shajib22}. Strong
lenses with time-variable background sources enable one-step measurements of cosmological distances from the time delays between multiple images,
allowing constraints on the cosmic expansion rate \citep[e.g.,][]{refsdal64,wong20,shajib23}. Conducting these studies with strongly lensed
supernovae is one of the scientific goals of our Highly Optimized Lensing Investigations of Supernovae, Microlensing Objects, and Kinematics of
Ellipticals and Spirals \citep[HOLISMOKES,][]{suyu20} program \citep[see also][for a review]{suyu23}.

Deep, wide-scale surveys are needed to identify statistically-significant samples of $\gtrsim 10^5$ strong lenses. Either or both imaging and
spectroscopic data sets can be relevant for this task, depending on the nature of the deflector and background source populations. For the 
galaxy-galaxy strong-lens category, the identification of spatially-resolved multiple images forming around a central galaxy is an efficient
selection technique which has been applied to several imaging surveys \citep[e.g.,][]{gavazzi14,marshall16,diehl17}. In the next years, the
{\it Euclid} \citep{laureijs11} and {\it Roman} telescopes \citep{green12}, and the Chinese Space Station Telescope will significantly expand
these imaging data sets in the optical and near-infrared from space. The {\it Rubin} Observatory Legacy Survey of Space and Time
\citep[LSST,][]{ivezic19} will cover similar wavelengths from the ground, and the Square Kilometer Array \citep[SKA,][]{mckean15} will open
a complementary window in the radio. Since the strong-lens discovery rate scales with both depth and spatial coverage, these next generation
surveys are expected to transform the field and to increase current sample of strong-lens candidates by at least two orders of magnitudes
\citep{collett15}. This nonetheless relies on highly-efficient, automated selection methods.

Machine learning techniques appeared in astronomy over the last decade, and supervised convolutional neural networks \citep[CNNs;][]{lecun98}
have since played increasing roles in image analysis problems. Besides galaxy morphological classification \citep[][]{dieleman15,walmsley22},
CNNs have also proven useful to estimate galaxy properties ranging from photometric redshifts \citep[e.g.][]{disanto18,schuldt21b} to
structural parameters \citep[e.g.,][]{tuccillo18,tohill21,li21b}. Given the possibility to simulate large samples of strong lenses with
highly-realistic morphologies for training, supervised CNNs have become state of the art for lens searches \citep{metcalf19}, and they have
been used for lens modeling \citep[e.g.,][]{hezaveh17,schuldt21a,schuldt22a,pearson21}. Other semi-supervised and unsupervised lens-finding
approaches are being developed \citep[e.g.,][]{cheng20,stein22}, but they do not yet offer a significant gain in classification accuracy.

Large samples of strong-lens candidates have been identified by applying supervised CNNs to existing surveys and by cleaning the network
outputs visually \citep[e.g.,][]{petrillo17,jacobs19b,huang20}. Recently, \citet{tran22} conducted the first systematic spectroscopic
confirmation of candidates selected through this process, finding a large majority of genuine strong lenses and a success rate of nearly
90\%. Human inspection is nonetheless key to reach this optimal purity. CNNs reach $>$99\% accuracy for balanced data sets of lenses and
nonlenses \citep[e.g.][]{lanusse18} but, since the fraction of strong lenses only represent up to 10$^{-3}$ of all galaxies per sky area,
applying these networks to real survey data result in samples dominated by false positives. Nearly 97\% of the candidates identified with
the PanSTARRS CNN from \citet{canameras20} were for instance discarded from the final list of high-quality candidates by visual inspection,
a rate comparable to other lens finders in the literature.

Low ratios of high-quality candidates over network recommendations imply long visual inspection stages. Moreover, \citet{rojas23} show that
grades attributed by groups of classifiers vary substantially, in particular for lenses with faint arcs and low Einstein radii, even when
restricting to ``expert classifiers''. Asking a single person to classify a sample of candidates multiple times also leads to substantial
scatter in the output grades, especially for non-obvious lenses \citep[see][]{shu22,rojas23}. Most classification biases can be minimized by
requesting multiple ($\gtrsim 5$--10), independent visual grades per candidate, but this further increases the need in human resources. The
larger number of galaxies in the next generation surveys will strongly complicate this process, even with crowdsourcing. By scaling the
performance of the PanSTARRS lens finder from \citep{canameras20} to deeper, higher-quality imaging, we expect for instance about 0.5 to 1
million CNN lens candidates to inspect over the {\it Rubin} LSST footprint. This clearly shows the need to further test CNN selections and
improve their recall with low contamination. Our subsequent strong-lens selection in Hyper-Suprime Cam imaging \citep{canameras21} already
showed an improvement in that regard.

In this work, we evaluate and compare systematically the performance of supervised neural networks in order to reduce human inspections for
future surveys. In general, test sets drawn from strong-lens simulations and user-dependent selections of nonlens galaxies can only roughly
mimick a classification experiment, given the variety of contaminants or image artefacts encountered in real survey data. In this study, we
build representative test sets directly from survey data to robustly evaluate the actual network performance and contamination rates. We use
multiband images from the Hyper-Suprime Cam Subaru Strategic Program \citep[HSC-SSP;][]{aihara18a} to train, validate and test our networks,
while taking advantage of previous searches for galaxy-scale strong lenses conducted in this survey with non-machine learning techniques
\citep{sonnenfeld18,sonnenfeld20,wong18,chan20,jaelani20}. HSC-SSP will also serve as testbed for the preparation of LSST with comparable
imaging quality, and depth only about 1~mag lower than the 10-year LSST stacks \citep[for the LSST baseline design,][]{ivezic19}\footnote{HSC
images are also processed with the pipeline that will be used for LSST \citep{juric17}.}. A forthcoming, complementary study (More et al., in
prep.) will compare machine learning assisted strong-lens finders trained by independent teams, using a range of simulation pipelines that are
not explored in this work, in order to gain precise insights on the relative network performance and selection functions.

The present paper is organized as follows. In Section~\ref{sec:method}, we introduce the overall procedure. The construction of ground-truth
data sets for training the CNNs is described in Section~\ref{ssec:train} and the test sets are defined in Section~\ref{ssec:test}. The various
network architectures are introduced in Section~\ref{sec:archi}. Section~\ref{sec:perf} presents the tests of several neural networks and
highlights the ingredients for reaching higher performance. Section~\ref{sec:discu} further discusses the stability of the network predictions
and Section~\ref{sec:conclu} summarizes the results. We adopt the flat concordant $\Lambda$CDM cosmology with $\Omega_{\rm M}=0.308$, and
$\Omega_\Lambda=1-\Omega_{\rm M}$ \citep{planck16}, and with $H_0=72$~km~s$^{-1}$~Mpc$^{-1}$ \citep{bonvin17}.

\section{Methodology}
\label{sec:method}

A comparison of algorithms for galaxy-scale lens searches was initiated by the {\it Euclid} consortium, using simulations from the Bologna
lens factory project \citep{metcalf19} based on the Millennium simulation \citep{lemson06}. This challenge demonstrated the higher performance
of CNNs compared to traditional algorithms, not only for the classification of single-band, {\it Euclid}-like images, but also for multi-band
data similar to ground-based surveys. Images for training and testing the networks were fully simulated, with the surface brightness of lens
and nonlens galaxies inferred from the properties of their host dark-matter halo and semi-analytic galaxy formation models. Even though disk,
bulge components, and spiral arms were also mocked up, the design of this challenge does not quantify the actual performance of lens-finding
algorithms on real data. More realistics test sets drawn from observed images and including the main populations of lens-like contaminants
(spirals, ring galaxies, groups) are needed to evaluate the ability of supervised neural networks in distinguishing strong lenses from the
broad variety of nonlens galaxies and image artefacts present in survey data. In this paper, we aim at filling in this gap by comparing the
performance of several deep learning strong-lens classifiers on real, multiband ground-based imaging observations including realistic number
and diversity of lens-like galaxies that are nonlenses.

We use data from the second public data release \citep[PDR2,][]{aihara19} of the HSC-SSP. This deep, multiband survey is conducted with the
wide-field HSC camera mounted on the 8.2\,m Subaru telescope and consists of three layers (Wide, Deep, and UltraDeep). We focus on the Wide layer
which aims at imaging 1400\,deg$^2$ in the five broadband filters $grizy$ with 5$\sigma$ point-source sensitivities of 26.8, 26.4, 26.2, 25.4,
and 24.7\,mag, respectively. The PDR2 observations taken until January 2018 cover nearly 800\,deg$^2$ in all bands down to limiting magnitudes of
26.6, 26.2, 26.2, 25.3, and 24.5\,mag in $grizy$, respectively, close to the survey specifications, with 300\,deg$^2$ having full-color full-depth
observations. The seeing distributions have median and quartile values in arcsec of $0.77^{+0.09}_{-0.08}$, $0.76^{+0.15}_{-0.11}$, $0.58^{+0.05}_{-0.05}$,
$0.68^{+0.08}_{-0.07}$, and $0.68^{+0.12}_{-0.09}$ in $g$, $r$, $i$, $z$, and $y$ bands, respectively.

\begin{figure*}
\centering
\includegraphics[width=.99\textwidth]{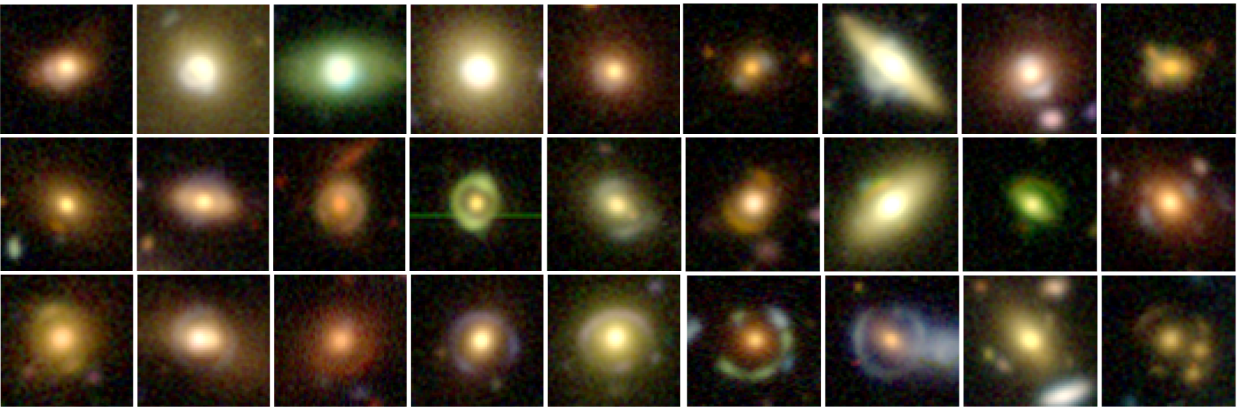}
\caption{Mocks in our baseline ground truth data set. The first, second, and third rows contain mocks with $0.75\arcsec<\theta_{\rm E}<1.30\arcsec$,
  $1.30\arcsec<\theta_{\rm E}<1.90\arcsec$, and $1.90\arcsec<\theta_{\rm E}<2.50\arcsec$, respectively. Cutouts have sizes of
  10\arcsec\,$\times$\,10\arcsec.}
\label{fig:mocks}
\end{figure*}

The tests in this paper are conducted on galaxies from HSC Wide PDR2 with at least one exposure in all five filters, including regions
that do not reach full depth in all filters. Experiments are mainly conducted with the $gri$ bands that have optimal depth, with $z$ band added
in some cases, and galaxies flagged with issues\footnote{Using the following flags in $grizy$ bands: {\tt cmodel\_flux\_flag, pixelflags\_edge,
  pixelflags\_interpolatedcenter, pixelflags\_saturatedcenter, pixelflags\_crcenter, pixelflags\_bad, sdsscentroid\_flag}} in HSC tables are
discarded. We focus on the subset of extended galaxies with Kron radius larger than 0.8\arcsec\ in the $i$ band and with $i$-band magnitudes
lower than 25~mag used by \citet{schuldt21b}, in order to limit the data volume while only excluding the faintest, most compact galaxies that
are unlikely to act as strong lenses. This provides a catalog of 62.5 million galaxies spanning a broad variety of morphological types. We note
that, while the survey strategy results in spatial variations in seeing and depth over the PDR2 images (see Sect.~\ref{sec:discu}), our parent
sample with $i$-band Kron radius $\geq$0.8\arcsec\ and $i<25$~mag is evenly distributed over the footprint and closely matches the average
properties listed in \citet{aihara19}. 

Rather than targeting all types of strong lens configurations, which would be challenging for an individual algorithm, we focus here on the
galaxy-galaxy systems which have a broad range of applications in astrophysics and cosmology. We intentionally avoid training and testing the
networks on systems with complex lens potentials such as those with multiple galaxies, and systems with a main, isolated deflector and strong
external shear from the lens environment. Lastly, the selection is aimed at optimizing the recall of systems with foreground luminous red galaxies
(LRGs) which have the highest lensing cross-section \citep{turner84}, and with any type of background source. We intend to find the main ingredients
to optimize the network contamination rates directly from the HSC images, without using strict pre-selections in color-color space, while also
minimizing the need for human validation of the network outputs.

The neural networks are trained, validated and tested on image cutouts from PDR2 with constant sizes of 60\,$\times$\,60 pixels
(10\arcsec\,$\times$\,10\arcsec), sufficient to cover the strong lensing features for galaxy-scale lenses with Einstein radii $\theta_{\rm E}<3$\arcsec.
The GAMA09H field -- which includes COSMOS -- is systematically discarded for training and validation. This field is reserved for a detailed comparison
of deep learning classifiers from several teams (More et al., in prep.), and also used to test the dependence of the network inference on variations
in seeing FWHM (see Sect.~\ref{sec:discu}).

\section{Data sets}
\label{sec:datasets}

\subsection{Ground truth data for training and validation}
\label{ssec:train}

We follow a supervised machine learning classification procedure, by training the neural networks on various sets of strong lens simulations and
nonlens galaxies. The construction of these sets of positive and negative examples are described below. The resulting ground truth data sets are
balanced, with 50\% positive and 50\% negative examples.

\subsubsection{Strong-lens simulations}

The simulations of galaxy-scale strong gravitational lenses were obtained with the pipeline described in \citet{schuldt21a,schuldt22a}. To produce
highly-realistic mocks capturing the properties of HSC images in the Wide layer, the pipeline paints lensed arcs on multiband HSC images of galaxies
acting as strong lens deflectors. This approach enables the inclusion of light for neighboring galaxies and accounts for the small-scale variations
in seeing and depth over the footprint. The foreground mass potentials were modeled using Singular Isothermal Ellipsoids (SIE) and, for each system,
only the SIE of the primary lens was considered. To assign a realistic SIE mass to each deflector, we focused on lens galaxies with robust
spectroscopic redshifts, $z_{\rm spec}$, and velocity dispersions, $v_{\rm disp}$, as described below. The SIE centroids, axis ratios, and position
angles were then inferred from the $i$-band light profiles, with random perturbations following the mass-to-light offsets measured in SLACS lenses
\citep{bolton08}. External shear was included in our simulations similarly as \citet{schuldt22a}, using a flat distribution in shear strength between
0 and 0.1 to cover plausible values in real lens systems, and using random shear position angles. The sample of lens LRGs with $z_{\rm spec}$ and
$v_{\rm disp}$ measurements was selected from the SDSS catalogs. After excluding the flagged QSOs, we collected all LRGs from the BOSS
\citep{abolfathi18} and eBOSS \citep{bautista18} catalogs with $\delta v_{\rm disp} < 100$~km~s$^{-1}$ to cover the broadest redshift range possible.
The resulting sample contains 50\,220 LRGs within the HSC Wide footprint with a redshift distribution peaking at $z \simeq 0.5$ and extending out to
$z \lesssim 1$, and a velocity dispersion distribution peaking at $v_{\rm disp} \simeq 200$~km~s$^{-1}$.

To include realistic source morphologies in our simulations we used high-resolution, high-SNR images of distant galaxies from the {\it Hubble} Ultra
Deep Field \citep[HUDF;][]{beckwith06} rather than simple parametric descriptions of the source light profiles. Using real sources accounts for the
diversity and complexity of high-redshift galaxies. We focused on the 1574 HUDF sources with spectroscopic redshift measurements from MUSE
\citep[${\tt CONFID} \geq 1$;][]{inami17}, resulting in a redshift distribution covering up to $z \simeq 6$, with two main peaks at $z \simeq 0.5$--1
and $z \simeq 3$--3.5. Before loading the {\it HST} exposures into the simulation pipeline, the neighbouring galaxies around the HUDF source with
measured $z_{\rm spec}$ were masked with SExtractor \citep{bertin96} as described in \citet{schuldt21a}. Color corrections were also applied to match
HST filter passbands to the HSC zeropoints.

The deflectors from the LRG sample were paired with random HUDF sources to satisfy specific criteria on the parameter distributions. The Einstein
radius, source color, and lens redshift distributions were controled during this stage to produce different data sets (see Sect.~\ref{ssec:testdata}).
A given lens LRG was included up to four times, with different sources and source positions, as we allowed rotations of the HSC cutouts by
$k \times \pi/2$, where $k=0$,1,2,3. For a given lens-source pair, the source was randomly placed in the source plane, over regions satisfying a lower
limit on the magnification factor $\mu$ of the central pixel, and then lensed with the {\tt GLEE} software \citep{suyu10,suyu12}. The lensed source was
convolved with the subsampled PSF model for the location of the lens released in PDR2\footnote{https://hsc-release.mtk.nao.ac.jp/doc/index.php/data-2/},
and scaled to the HSC pixel size and to the HSC photometric zeropoints. Lensed images were finally coadded with the lens HSC cutout. To ensure that all
simulations include bright, well-detected arcs, we required that the brightest pixel over the lensed arcs exceeds the background noise level over the
lens LRG cutout by a factor SNR$_{\rm bkg, min}$, either in $g$ or $i$ band depending on the source color. While most lensed images are necessarily blended
with the lens galaxies, we further required that the brightest pixel over the lensed arcs has a flux higher than the lens galaxy at that position by a
factor $R_{\rm sr/ls, min}$. For each lens-source pair, the source was randomly moved in the source plane until the lensed images satisfy these conditions;
if the conditions were not satified after 20 iterations, then the source brightness was artificially boosted by 1\,mag in each band. The procedure was
repeated until reaching the maximal magnitude boost of 5\,mag.

Our baseline data set includes 43\,750 mock lenses with flat Einstein radius distribution over the range 0.75\arcsec$-$2.5\arcsec, and with lensed
images having $\mu \geq 5$, SNR$_{\rm bkg, min} = 5$, and $R_{\rm sr/ls, min} = 1.5$ (see Fig.~\ref{fig:mocks}). This lower limit on $\theta_{\rm E}$
approaches the median seeing FWHM in $g$ and $r$ bands, and helps obtain multiple images that meet our brightness and deblending criteria. The
matching of lens-source pairs applied stronger weights on lens LRGs at $z_{\rm d}>0.7$ in order to increase the relative fraction of fainter and
redder lenses in the data set. Similarly, the number of red sources was boosted to increase the fraction of red lensed arcs by a factor two compared
to the original HUDF sample. Other sets of simulations tested in Sect.~\ref{ssec:testdata} use alternative parameter distributions, and different
values of $\mu$ and SNR$_{\rm bkg, min}$, and they contain between 30\,000 and 45\,000 mocks.

\subsubsection{Selection of nonlenses}

The samples of nonlens galaxies used in our various ground truth data sets were selected from the parent sample of galaxies with $i$-band Kron
radius $\geq$0.8\arcsec\ to match the restriction on the lens sample. In order to investigate the classification performance as a function of the
morphological type of galaxies included in the training sets, we extracted specific samples of galaxies using publications in the literature. We
focused on the galaxy types forming the majority of nonlens contaminants, in order to help the neural networks learn models that are able to
identify the strong lensing features while excluding the broad variety of nonlens galaxies. In the following, we give specific details on each of
these samples.

The extended arms of spiral galaxies can closely resemble the radial arcs formed by strongly lensed galaxies, especially for low inclination
angles. Given that spiral arms predominantly contain young, blue stellar populations, akin to the colors of high-redshift lensed galaxies, the
two types can present very similar morphologies ground-based seeing-limited optical images. We based our selection of spirals on the catalog
of \citet{tadaki20}. This study visually identifies 1447 S-spirals and 1382 Z-spirals from HSC PDR2 images to train a CNN and find a larger
sample of nearly 80\,000 spirals with $i<20$~mag over 320~deg$^2$ in the Wide layer. Given the construction of their training set, the selection
of \citet{tadaki20} is mainly sensitive to large, well-resolved spiral arms with clear winding direction. We tried various cuts on the galaxy
sizes, finding that $i$-band Kron radii $\leq$2\arcsec\ is an optimal cut to obtain a sufficient number of 40\,000 spirals while ensuring that
spiral arms fall at 1$-$3\arcsec\ from the galaxy centroids and within our 10\arcsec\,$\times$\,10\arcsec\ cutouts.

The networks in our study were trained to identify the importance of strong lensing features such as extended arcs or multiple images by including
large fractions of isolated LRGs in the data sets. To focus on the brightest, massive LRGs that are likely to act as strong lenses, we selected
LRGs from the same parent sample as in our simulations. Moreover, groups of bright galaxies within 5--10\arcsec\ in projection that mimick the
distribution of multiple images are frequently misclassified as strong lenses. We constructed a sample of compact groups using the catalog of
groups and clusters from \citet{wen12} based on SDSS-III. The richest structures were selected by setting the number of galaxies within a radius
of $r_{\rm 200}$ to $N_{\rm 200}>10$, and by requiring at least three bright galaxies with $r_{\rm Kron}<23$ within 10\arcsec. The cutouts were then
centered on the HSC galaxy closest to the position given by \citet{wen12}. Finally, random nonlens galaxies were selected from the parent sample
with $i$-band Kron radius $\geq$0.8\arcsec, after excluding all confirmed and candidates lenses from the literature, using our compilation up to
December 2022 (see \citet{canameras21} and references therein). Various cuts on the $r$-band Kron magnitudes were tested, including the criteria
$r<23$~mag that covers the majority of LRG lens galaxies, but we obtained better performance for random nonlenses down to the limiting magnitude
of HSC Wide.

Other classes such as edge-on galaxies, rings, and mergers can possibly confuse the neural networks \citep[see][]{rojas22} but were not directly
included in the nonlens set. While citizen science projects based on SDSS \citep{willett13}, HST \citep{willett17}, and DECaLS \citep{walmsley22}
include such morphological types in their classification, they overlap only partially with the HSC footprint and do not provide the $\gtrsim$10$^3$
examples required to further tune our data sets. Studies based on unsupervised machine learning algorithms \citep[e.g.,][]{hocking18,martin20,cheng21}
are allowing efficient separation of early- and late-type galaxies, but their ability to identify pure sample of rare galaxy types needs further
confirmation. In the future, outlier detection could become an alternative \citep[e.g.,][]{margalef20}.

The nonlenses in our baseline ground truth data set include 33\% spirals, 27\% LRGs, 6\% groups, and 33\% random galaxies. Other alternative data
sets tested in Sect.~\ref{ssec:testdata} either use only one of these morphological types, or vary their relative proportion. Similar to the LRGs
in the lens simulations, nonlens galaxies cover random position over the entire HSC PDR2 footprint. This ensures that galaxies in our training set
sample representative seeing FWHM values and image depth, which is particularly important given that only about 40\% of the area we consider
reaches nominal depth in all five bands.

\subsection{Content of the test sets}
\label{ssec:test}

The performance of our classification networks are evaluated on two specific test sets, which are also drawn from the input sample of real
galaxies in HSC PDR2 with $i$-band Kron radius $\geq$0.8\arcsec. While the discussion resulting from this analysis is directly related to the
construction of these test sets, we expect the results to be easily transferrable to other HSC data releases and to external surveys with
comparable image depth and quality.

\subsubsection{Strong lenses from the literature}

Spectroscopically-confirmed or high-quality candidate galaxy-scale strong lenses from the Survey of Gravitationally-lensed Objects in HSC
Imaging (SuGOHI) were used to test the neural network recall \citep{sonnenfeld18,sonnenfeld20,wong18,chan20,jaelani20}. These systems have
been previously discovered with multiband imaging from the Wide layer using non-machine learning selection techniques followed by visual
inspection from experts. As their selection relies on the combination of various techniques -- such as searches for spectral lines in blended
spectra, lens light subtraction and lens modeling, or crowdsourcing -- these systems form a representative subset of the overall population
of detectable strong lenses in HSC. We selected the highest-quality SuGOHI lenses classified as grade A or B according to the criteria listed
in \citet{sonnenfeld18}. Given our focus on galaxy-scale lenses, we visually excluded 27 systems with image separations $\gg$3\arcsec\ typical
of group-scale lenses, or with lensed arcs significantly perturbed by nearby galaxies or large-scale mass components. We checked that none
of the remaining lenses contains background quasars, and we restricted to lens galaxies with $i$-band Kron radius above 0.8\arcsec\ to match
our overall search sample. This results in a sizeable set of 189 test lenses.

Some of these SuGOHI lenses and lens candidates were found in earlier data releases covering smaller areas than PDR2, but the images
originally used for discovery also cover $gri$ bands, with depth comparable to PDR2 \citep[see][]{aihara18b,aihara19}. In terms of angular
separation between the lens center and multiple images, the various SuGOHI classification algorithms apply lens light subtraction prior to
the arc identification and lens modeling steps, which presumably helps identify more compact systems \citep[see][]{sonnenfeld18}. The subset
with detailed lensing models nonetheless have Einstein radii in the range 0.80--1.80\arcsec, which is representative of the distribution over
the entire sample peaking at $\theta_{\rm E} \simeq 1.2$\arcsec--1.3\arcsec. This indicates that the 189 test lenses have both well-deblended
lens and source components and sufficiently high S/N in $gri$ images from the PDR2 Wide layer, and that all should be recovered via deep-learning
classification of raw images. In contrast, additional lenses and lens candidates that are not detected and spatially-resolved in PDR2 $gri$
cutouts \citep[e.g., from SDSS fiber spectra,][]{bolton08,brownstein12,shu16} were discarded for testing.

\subsubsection{Nonlenses in the COSMOS field}

We collected a large sample of nonlens galaxies in the COSMOS field to quantify the ability of our networks to exclude the broad variety of
contaminants, and to obtain the most realistic false positive rate estimates for a real classification setup. By focusing on the well-studied
COSMOS field, we can firmly exclude all strong lenses and conduct these estimates automatically. Nonlenses were selected from our parent
sample with Kron radius larger than 0.8\arcsec\ and without flagged cutouts. We note that these flags exclude galaxies with unreliable
photometry, but do not exclude cutouts with partial coverage in one or several bands, with diffrations spikes, or other artefacts. Moreover,
\citet{aihara19} show that a few artefacts remain in the coadded frames of the Wide layer, such as compact artefacts near static sources, and
artefacts located in regions with only one or two exposures in PDR2. These were intentionally kept in our test set. All confirmed lenses and
lens candidates were excluded using the MasterLens database\footnote{http://admin.masterlens.org}, \citet{faure08}, \citet{pourrahmani18},
and SuGOHI papers, leaving 70,910 unique nonlens galaxies.

To match our overall approach, we focused on the Wide layer and ignored COSMOS images from Deep and UDeep layers. In PDR2, the COSMOS field
is observed to full-depth in all filters, which is not the case for all HSC fields included in our parent sample. We probed differences in
image quality between COSMOS and the overall footprint by plotting distributions of the number of input frames per coadd and of the seeing
FWHM in $gri$ bands, for the 70,910 nonlenses in COSMOS and a random subset of our parent sample. The distributions of number of frames per
band roughly match for both samples. Only the tail of $\lesssim 10$\% galaxies with $\leq$3 frames per stack disappears in the COSMOS field.
The distributions of seeing FWHMs differ more strongly, since the small COSMOS field was observed with atmospheric conditions closest to the
survey specifications. Fortunately, in COSMOS, the median seeing FWHMs in $gri$ bands closely match the values over the HSC footprint listed
in Sect.~\ref{sec:method}, with only a smaller scatter. The quality of PDR2 images in our COSMOS test set are therefore roughly representative
of the overall HSC Wide footprint.

\section{The network architectures}
\label{sec:archi}

State of the art CNN and ResNet architectures have been tested, starting with baseline architectures and exploring variations around this baseline.
In this section, we describe these architectures together with alternative group-equivariant neural networks aimed at improving the stability with
respect to image rotations. More advanced supervised machine learning approaches are increasingly used in astronomy. For instance,
\citet[][]{thuruthipilly21} implemented self-attention-based architectures \citep[Transformers,][]{vaswani17} for lens searches using simulated data
from the Bologna lens challenge. Such modern neural network models remain nonetheless prone to the class imbalance and domain adaptation issues
affecting classical CNNs \citep[see, e.g.,][]{thuruthipilly22} and further work in these directions are postponed to future studies.

\subsection{Baseline convolutional neural network}

We used CNN architectures inspired from AlexNet \citep{krizhevsky12}, with a baseline architecture comprising three convolutional layers and three
fully-connected (FC) hidden layers, before the single-neuron output layer resulting in the network prediction. Rectified Linear Unit
\citep[ReLU,][]{nair10} activation functions are placed between each of these layers to add nonlinearity into the network, and sigmoid activation
is applied to the last layer. The convolutional layers have kernels with sizes 11\,$\times$\,11, 7\,$\times$\,7, and 3\,$\times$\,3, and they have
32, 64, and 128 feature maps, respectively. The feature maps in the convolutional layers are downsampled to improve invariance with respect to
translation of morphological features across the input images. To that end, we used a max-pooling layer \citep{ranzato07} with 2\,$\times$\,2
kernels and a stride of 2 after each of the first two convolutional layers. The FC layers have 50, 30, and five neurons each, and dropout
regularization \citep{srivastava14} with a dropout rate of 0.5 is applied before each of these FC layers.

\subsection{Baseline residual neural network}

Deeper networks were trained to characterize their ability to learn the small-scale features in multiband images and to quantify their overall
classification performance. We used residuals networks \citep[ResNet,][]{he16}, a specific type of CNNs implementing residual blocks to help
train much deeper architectures without facing the problem of vanishing gradients during back-propagation. Such ResNet contain multiple building
blocks, also called preactivated bottleneck residual units, separated by shortcut connections. Our baseline ResNet is inspired from the ResNet18
architecture while the deeper, standard residual networks ResNet34, ResNet50, and ResNet101 did not improve the classification accuracies
substantially for the small image cutout we considered. After the input images, the network contains a first convolutional layer with a
3\,$\times$\,3 kernel and 64 features maps followed by batch normalization \citep{ioffe15}. We implement eight blocks comprising two convolutional
layers with kernels of 3\,$\times$\,3 pixels, batch normalization and nonlinear ReLU activations. The blocks are grouped by two, with 64, 128,
256, and 512 feature maps per group, and strides of 1, 2, 2, and 2, respectively. Using larger kernels over the convolutional layers did not allow
extraction of the small-scale strong lens features and we therefore kept 3\,$\times$\,3 kernels. An average pooling layer with a 6\,$\times$\,6
kernel is used before flattening, and its output is passed to a FC hidden layer with 16 neurons and ReLU activation. The last layer contains a
single neuron with sigmoid activation.

\subsection{Group-equivariant neural network}

While standard CNN and ResNet architectures account for the spatial correlations in the images, they do not ensure that network predictions are
invariant to rotations and reflections. Several classification tasks have proven to benefit from neural network architectures able to directly learn
equivariant representations with a limited number of trainable parameters. This is the case for the separation of radio galaxies into Fanaroff Riley
types I and II, which suffers from the scarcity of labelled data and the difficulty to produce realistic simulations \citep[e.g.,][]{scaife21}. To
exploit the symmetries inherent to lens finding, we followed \citet{schaefer18} and \citet{scaife21} by testing group-equivariant neural networks
(G-CNNs). We used the G-CNN from \citet{cohen16} implemented in the GrouPy python library\footnote{https://github.com/tscohen/GrouPy} which achieved
excellent performance on the MNIST and CIFAR10 data sets.

Convolutional layers in CNNs ensure equivariance to the group of 2D translations by integers. The G-CNN architecture generalizes these properties and
exploits symmetries under other groups of transformations with specific, group-equivariant convolutional layers. These layers involve multiple kernels,
which are obtained by applying the transformations of the group under consideration to a single kernel. In our case, we impose that our networks learn
features equivariant to the ``p4m'' group of translations, mirror reflections, and rotations by $k \times \pi/2$ degrees. We build our architecture
using the standard G-CNN from \citet{cohen16} with a first classical convolutional layer followed by three p4 convolutional layers and two
fully-connected layers, with ReLU activations and sigmoid activation on the last one. After the second and fourth layers, two layers are inserted to
apply max-pooling over image rotations with 2\,$\times$\,2 kernels and a stride of 2. 

\begin{figure*}
\centering
\includegraphics[width=.49\textwidth]{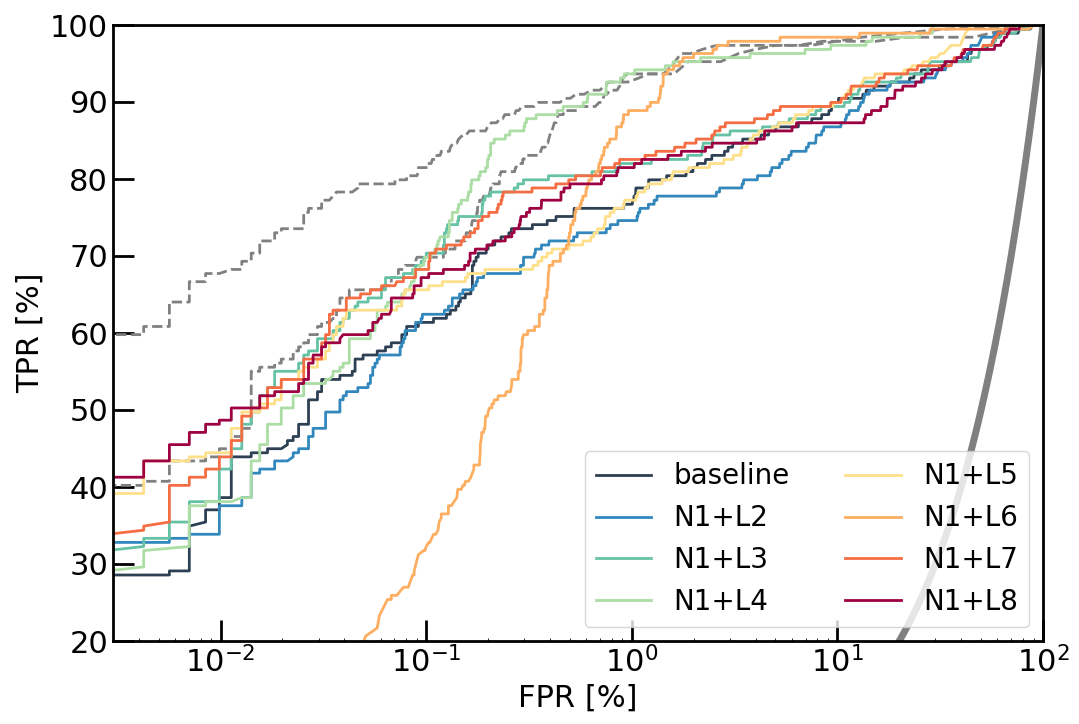}
\includegraphics[width=.49\textwidth]{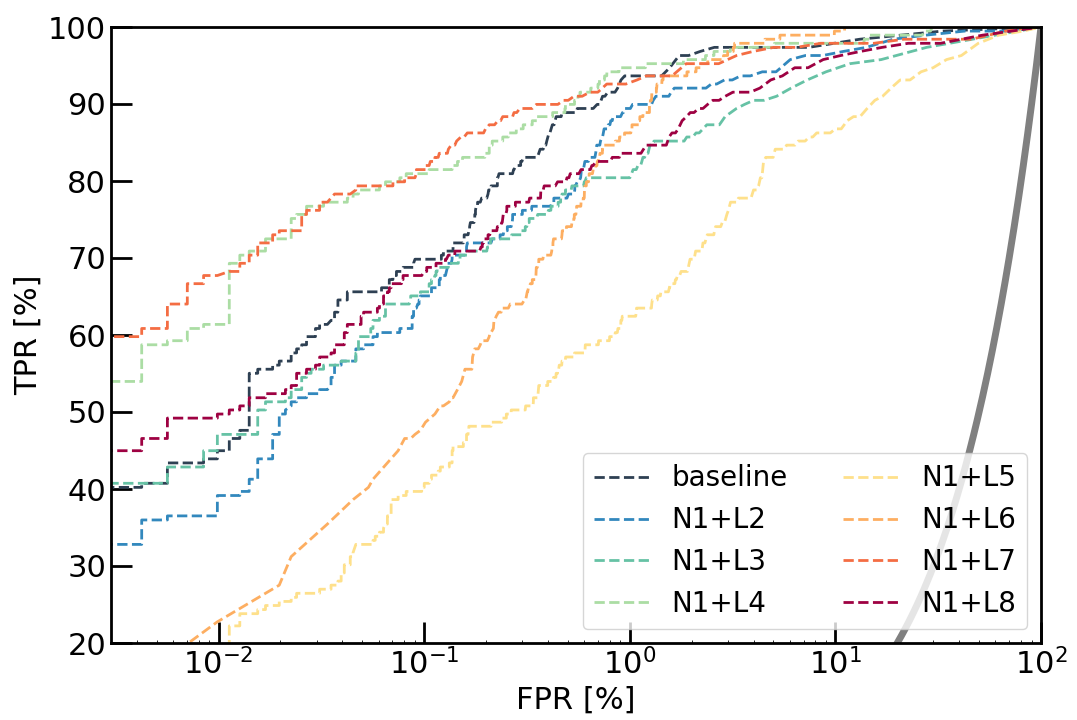}

\includegraphics[width=.49\textwidth]{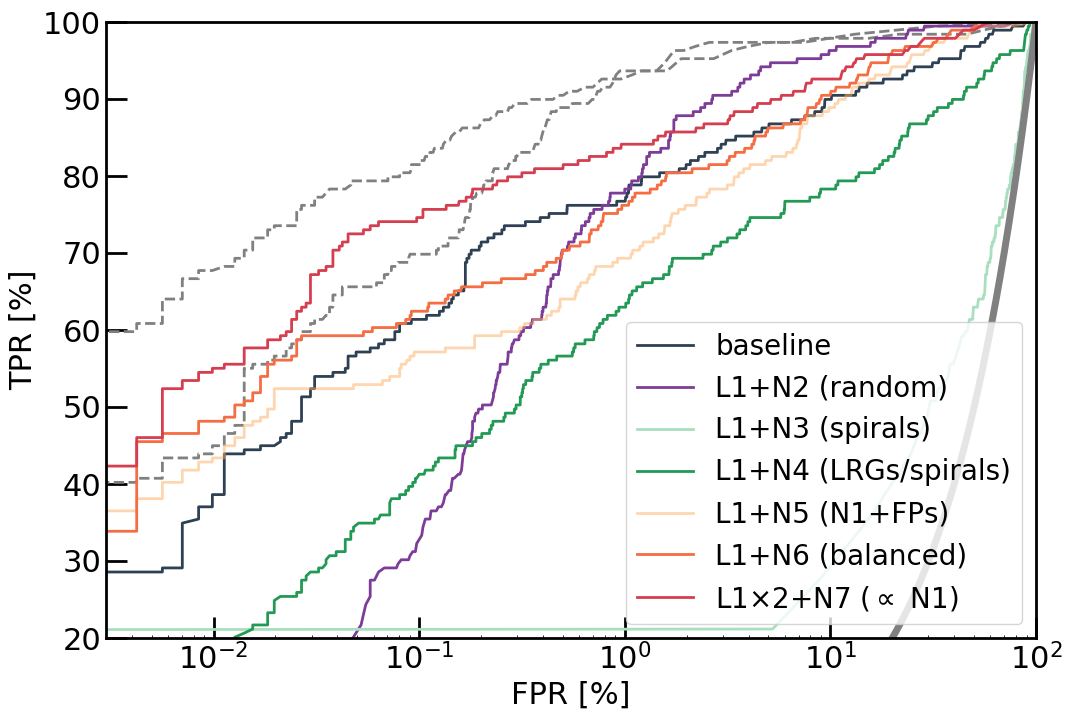}
\includegraphics[width=.49\textwidth]{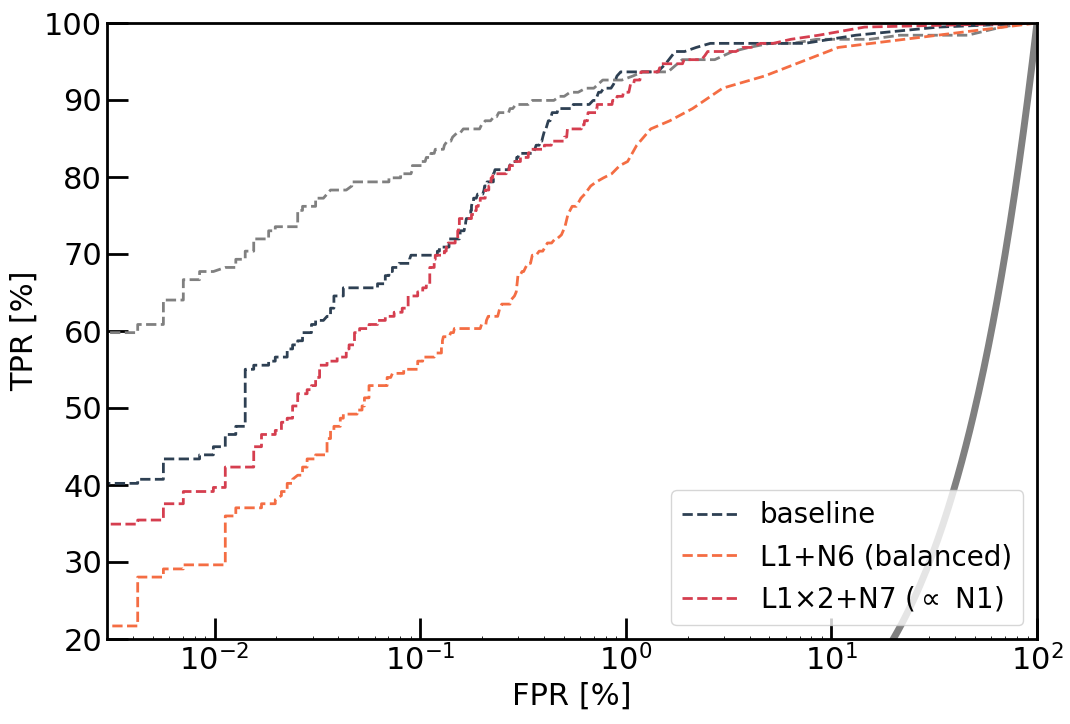}
\caption{Influence of the training data set for our baseline CNN (left, solid lines) and ResNet (right, dashed lines). We only vary the set of
  positive (top) and negative (bottom) examples. Networks trained on the baseline data sets (N1+L1) are plotted in dark blue. Two of the best
  networks from the upper-right panel, the baseline ResNet (N1+L1) and the ResNet from C21 (N1+L7), are shown as dashed grey lines in all panels
  for reference. The thick grey curve corresponds to a random classifier. Optimal performance are obtained for ground-truth data sets comprising
  mock lenses with bright, deblended multiple images, and large fractions of typical nonlens contaminants. The AUROC, TPR$_{\rm 0}$ and TPR$_{\rm 10}$
  tend to be higher for the ResNet, except for data sets containing limited numbers of tricky nonlens galaxies.}
\label{fig:dataset}
\end{figure*}

\begin{table}
\caption{Performance for various training data sets.}
\centering
\begin{tiny}
\begin{tabular}{lcccccc}
\hline
\hline \\[-0.5em]
 & \multicolumn{3}{c}{CNN} & \multicolumn{3}{c}{ResNet} \\
\cmidrule(lr){2-4}\cmidrule(lr){5-7}
Data set & AUROC & TPR$_{\rm 0}$ & TPR$_{\rm 10}$ & AUROC & TPR$_{\rm 0}$ & TPR$_{\rm 10}$ \\[+0.5em]
\hline \\[-0.5em]
baseline & 0.9557 & 0.0 & 44.4 & 0.9913 & 36.0 & 55.0 \\
(L1+N1) \\[+0.5em]
\hline \\[-0.8em]
\multicolumn{3}{l}{{\bf N1 + ...}} \\[+0.5em]
L2 & 0.9527 & 7.4 & 41.8 & 0.9860 & 26.5 & 41.3 \\
L3 & 0.9581 & 0.0 & 49.8 & 0.9728 & 27.5 & 47.1 \\
L4 & 0.9919 & 0.0 & 43.4 & 0.9941 & 42.9 & {\bf 70.9} \\
L5 & 0.9686 & 8.5 & 49.7 & 0.9502 & 10.1 & 23.8 \\
L6 & 0.9891 & 0.0 & 0.0 & {\bf 0.9948} & 0.0 & 0.0 \\
L7 & 0.9624 & 0.0 & 50.3 & 0.9870 & {\bf 49.2} & 70.4 \\
L8 & 0.9531 & 0.0 & 50.3 & 0.9799 & 38.6 & 51.9 \\ 
\hline \\[-0.8em]
\multicolumn{3}{l}{{\bf L1 + ...}} \\[+0.5em]
N2 & 0.9831 & 0.0 & 0.0 & \dots & \dots & \dots \\
N3 & 0.6141 & 0.0 & 0.0 & \dots & \dots & \dots \\
N4 & 0.8984 & 0.0 & 20.6 & \dots & \dots & \dots \\
N5 & 0.9610 & 32.8 & 47.6 & \dots & \dots & \dots \\
N6 & 0.9704 & 31.2 & 50.8 & 0.9762 & 19.1 & 37.0 \\
N7 & 0.9763 & 23.8 & 57.7 & 0.9929 & 30.2 & 42.3 \\
\hline
\end{tabular}
\tablefoot{The data sets are described in detail in Sect.~\ref{ssec:testdata}. Architectures were kept fixed to the baseline CNN and ResNet. The 
  ResNets trained on sets L1+N2, L1+N3, L1+N4, and L1+N5 are discarded since none of these networks reached acceptable performance
  (${\rm AUROC \gtrsim 0.8}$).}
\label{tab:dataset}
\end{tiny}
\end{table}

\section{Tests on the classification performance}
\label{sec:perf}

Multiple networks were trained with the various ground-truth data sets, architectures, and data pre-processing described below, using data set
splits of 80\% for training and 20\% for validation. We randomly initialized the network weights and trained the networks using mini-batch stochastic
gradient descent with 128 images per batch. We minimized the binary cross-entropy loss computed over each batch between the ground truth and predicted
labels. This loss function is standard for binary classification problems and enables us to penalize robust and incorrect predictions. We used a learning
rate of 0.0006, a weight decay of 0.001, and a momentum fixed to 0.9. Each network was trained over 300 epochs and the final model was saved at the
``best'' epoch corresponding to the lowest binary cross-entropy loss in the validation set. Hyperparameters were only modified for networks that either
showed plateaus in their training loss curves, large generalization gaps at the best epoch, or significant overfitting. In these cases, hyperparameters
were optimized via a grid search, by varying the learning rate over the range [0.0001, 0.1] and the weight decay over [0.00001, 0.01], both in steps of
factor 10, while keeping momentum fixed to 0.9. Networks showing no improvement after tuning these hyperparameter were discarded from the analysis.

\subsection{Definition of metrics}

The networks were compared based on the standard metrics used in binary classification experiments. First, the Receiver Operating Characteristic
(ROC) curves were computed using the following definitions of the true positive rate (TPR or recall) and false positive rate (FPR or contamination),
\begin{equation}
\rm TPR = \frac{TP}{TP+FN};\ FPR = \frac{FP}{FP+TN}
\end{equation}
by varying the score thresholds between {\tt 0} (nonlens, or negative) and {\tt 1} (lens, or positive). TP, FP, TN, and FN refer to the number of true
positives, false positives, true negatives, and false negatives, respectively. This allowed us to infer the area under the ROC (AUROC) by computing the
integral, in order to assess the networks, especially those approaching the ideal AUROC value of 1. Second, TPR$_{\rm 0}$ and TPR$_{\rm 10}$, which are
the highest TPRs for a number of false positives of 0 and 10 in the ROC curve, respectively, were derived to gauge the contamination of each network.
To help assess differences between panels, the ROC curves presented in the different figures systematically show two of our best networks, the baseline
ResNet and the ResNet from \citet[][hereafter \citetalias{canameras21}]{canameras21}. We focused our discussion on networks with excellent performance
in terms of AUROC and we relied on TPR$_{\rm 0}$ and TPR$_{\rm 10}$ to identify the networks with lowest contamination.

\subsection{Ground truth data set}
\label{ssec:testdata}

Like any supervised machine learning algorithm, our strong lens classification networks necessarily depend on the properties of galaxies included
in the ground truth data set. In this section, we determine to what extent the overall performance of our baseline CNN and ResNet vary as a function
of the arbitrary construction of the data sets. We first tested the performance of networks trained on the baseline set of negative examples (N1)
and various sets of realistic mocks constructed with our simulation pipeline, and we then tested the impact of fixing the baseline set of positive
examples (L1) and constructing the set of nonlenses in different ways.

{\noindent {\bf Samples of mocks lenses.}} The baseline set of lens simulations (L1) described in Sect.~\ref{ssec:train} contain bright arcs with 
$\mu \geq 5$, SNR$_{\rm bkg, min} = 5$, and $R_{\rm sr/ls, min} = 1.5$ either in $g$ or $i$ band, a flat Einstein radius distribution over the range
0.75\arcsec$-$2.5\arcsec, and external shear. Specific weights were applied as a function of the source color so that the overall $(V-i)$
distribution peaks at $\simeq$0.2, as the input HUDF sample, but with a fraction of red sources with $(V-i) \simeq 1$--2.5 increased by a factor
two. Apart from the modifications specified below, other sets are simulated as the baseline. {\bf Set\,L2} is identical to set L1, but lowers the
number of small-separation systems by restricting to 1.0\arcsec$-$2.5\arcsec. {\bf Set\,L3} moves the flat $\theta_{\rm E}$ distribution to the range
0.75\arcsec$-$2.0\arcsec. {\bf Set\,L4} uses a natural Einstein radius distribution decreasing from 4200 mocks in the lowest bin
$0.75\arcsec < \theta_{\rm E} < 0.80\arcsec$ to 150 mocks in the highest bin $2.45\arcsec < \theta_{\rm E} < 2.50\arcsec$. {\bf Set\,L5} boosts the
source brightness more progressively than L1, at the expense of a larger computational time, and only requests $R_{\rm sr/ls, min} = 1.0$. The resulting
mocks in set L5 ressemble the baseline mocks but with fainter lensed arcs, closer to the lower limits in source brightness defined by the parameters
SNR$_{\rm bkg, min}$, and $R_{\rm sr/ls, min}$. {\bf Set\,L6} imposes criteria on the image configurations. In general, restricting the source positions to
high-magnification regions closer to the caustic curves in the source plane results in larger numbers of quads and complete Einstein ring. Set L6 was
constructed similarly to L1 while discarding the threshold on $\mu$, and checking explicitly whether the lensed sources are doubly- or quadruply-imaged
to mitigate for this effect and to obtain a balanced set of image configurations. {\bf Set\,L7} follows the construction of the baseline mocks but
without boosting the fraction of red HUDF sources. This is the data set used by \citetalias{canameras21}. {\bf Set\,L8} simplifies the association of
lens-source pairs by discarding the boost of red sources and high-redshift lenses, and computes the brightness thresholds exclusively in the $g$ band.
Finally, other simulations were tested such as mocks with a flat lens redshift distribution over the range $0.2 < z_{\rm d} < 0.8$, but these sets are
not discussed below given their much lower AUROC and recall at zero contamination.

{\noindent {\bf Selection of nonlenses.}} The baseline set of nonlenses (N1) contains 33\% spiral galaxies, 27\% LRGs, 6\% compact groups, and 33\%
random galaxies. We tested various sets of nonlenses differing from set N1 by selecting galaxies from the parent samples introduced in
Sect.~\ref{ssec:train}. Sets N2, N3, and N4 include a single type of nonlens galaxies. {\bf Set\,N2} contains random galaxies without $r$-band
magnitude cuts, {\bf Set\,N3} only has spiral galaxies, and {\bf Set\,N4} joins spirals together with LRGs. {\bf Set\,N5} builds on set N1 by adding
nearly 5000 false positives from the visual inspection campaign from \citetalias{canameras21}. {\bf Set\,N6} groups the same morphological classes
as N5 but in different proportions to improve the class balance. Overall, set N6 includes 25\% spirals, 24\% LRGs, 12\% compact groups, 24\% random
galaxies, and 19\% nonlenses from previous networks. {\bf Set\,N7} contains twice the number of galaxies as the baseline, in the same proportions as
set N1. The network trained on N7 uses a duplicated version of set L1 as positive examples, in order to investigate the change in performance with a
larger data sets with larger number of nonlens galaxies.

The results are summarized in Fig.~\ref{fig:dataset} and Table~\ref{tab:dataset}. Despite our effort in restricting these tests to realistic sets of
positive and negative examples, we find that our neural networks are highly sensitive to choices in the construction of the ground truth data set.
When changing the sets of mocks, we obtain similar variations in AUROC for the CNN and ResNet architectures, but the TPR$_{\rm 0}$ and TPR$_{\rm 10}$
values are more stable for the CNN and, except TPR$_{\rm 10}$ for the networks trained on L5+N1, the recall at low contamination is systematically
higher for the ResNet. Except for L6+N1, the no-contamination recall TPR$_{\rm 0}$ remains $\gtrsim$10\% for the ResNet, reaching TPR$_{\rm 0}=42.9$\% 
for L4+N1 and TPR$_{\rm 0}=49.2$\% for the most restrictive ResNet from \citetalias{canameras21} trained on L7+N1. When varying the sets of nonlenses,
the metrics show that using random galaxies over the footprint L1+N2 results in suboptimal performance, with larger contamination rates and TPR$_{\rm 0}$
and TPR$_{\rm 10}$ both equal to zero for the CNN. This conclusion holds for all magnitude cuts that have been applied to the set of random nonlenses.
Neither the baseline CNN nor the alternative CNN architectures we tested in the following section allowed us to boost the performance without
fine-tuning the ratio of galaxy types used as negative examples. In addition, using only spiral galaxies as nonlenses in L1+N3 does not perform well
on the larger diversity of populations included in our test sets resulting in the lowest AUROC of Table~\ref{tab:dataset}. As previously discussed in
\citet{canameras20}, the performance increase substantially when jointly boosting the fraction of usual contaminants. The data sets N1, N6, and N7
constructed in such ways are resulting in the highest low-contamination recall for the CNN, and these are the only sets providing good AUROC with
the ResNet and included in Fig.~\ref{fig:dataset}. Over these tests, the best performance are obtained for the largest set L1+N7, which is also the
data set providing the most consistent results between the CNN and the ResNet.

\begin{figure*}
\centering
\includegraphics[width=.49\textwidth]{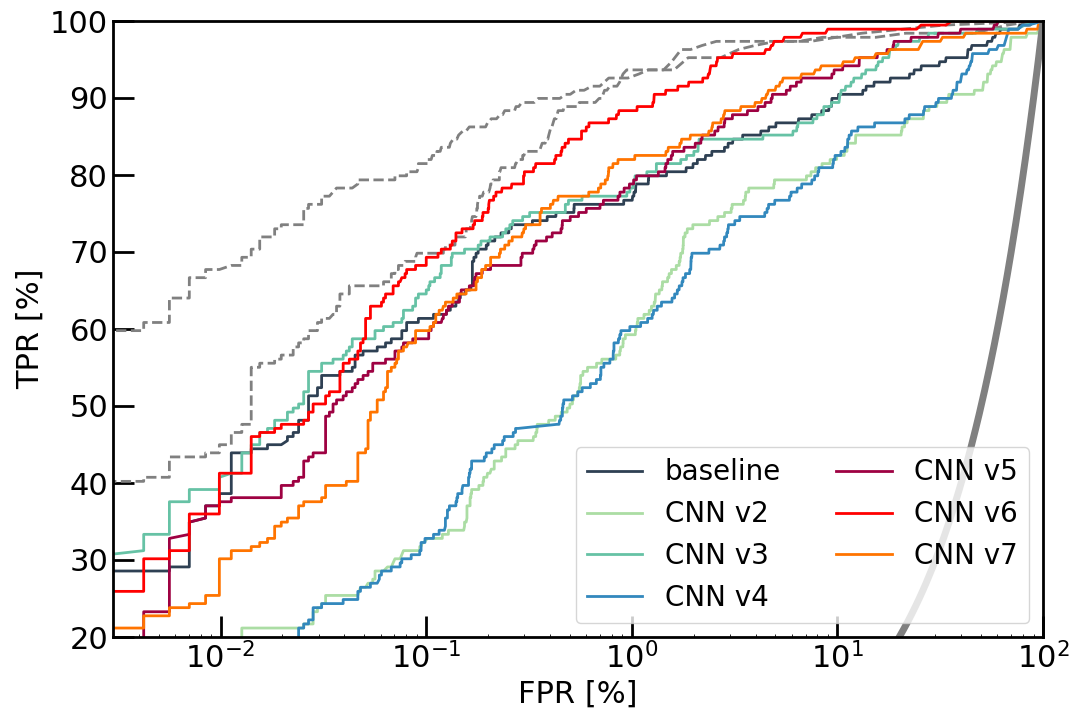}
\includegraphics[width=.49\textwidth]{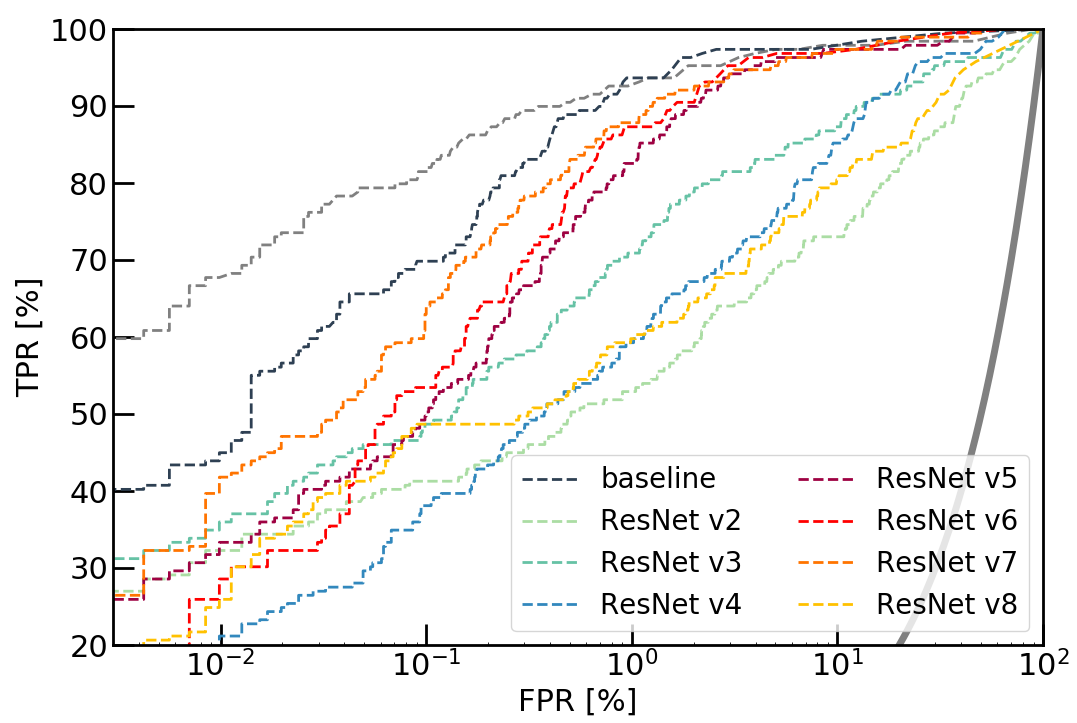}
\caption{Influence of the network architecture, for our baseline data set and various CNNs (left) and ResNets (right). The baseline architectures are
  in dark blue. The group-equivariant network architectures G-CNNs adapted from \citet{cohen16} are plotted in Appendix. For reference, the dashed
  grey lines show two good networks (the baseline ResNet and the ResNet from C21). The thick grey curve corresponds to a random classifier. The best
  performance is obtained for the baseline ResNet.}
\label{fig:archi}
\end{figure*}

\begin{table}
\caption{Performance for various CNN and ResNet architectures.}
\centering
\begin{tiny}
\begin{tabular}{lccc}
\hline
\hline \\[-0.5em]
Architecture & AUROC & TPR$_{\rm 0}$ & TPR$_{\rm 10}$ \\[+0.5em]
\hline \\[-0.5em]
baseline CNN & 0.9557 & 0.0 & 44.4 \\
CNN v2 & 0.9115 & 10.6 & 21.2 \\
CNN v3 & 0.9694 & 0.0 & 45.0 \\
CNN v4 & 0.9242 & 0.0 & 13.2 \\
CNN v5 & 0.9772 & 1.6 & 38.1 \\
CNN v6 & 0.9927 & 10.1 & 46.0 \\
CNN v7 & 0.9704 & 7.4 & 31.8 \\
G-CNN v1 & 0.9705 & 13.8 & 51.9 \\
G-CNN v2 & 0.9865 & 7.9 & 41.8 \\
G-CNN v3 & 0.9883 & 15.9 & 48.2 \\
\hline \\[-0.5em]
baseline ResNet & {\bf 0.9913} & {\bf 36.0} & {\bf 55.0} \\
ResNet v2 & 0.8939 & 21.2 & 34.4 \\
ResNet v3 & 0.9472 & 28.0 & 37.0 \\
ResNet v4 & 0.9499 & 10.1 & 22.8 \\
ResNet v5 & 0.9857 & 16.9 & 34.4 \\
ResNet v6 & 0.9885 & 12.2 & 30.2 \\
ResNet v7 & 0.9884 & 19.6 & 43.9 \\ 
ResNet v8 & 0.9199 & 15.9 & 31.8 \\
\hline
\end{tabular}
\tablefoot{Architectures correspond to variations of the baseline CNN, G-CNN and ResNet introduced in Sect.~\ref{sec:archi}. Further details are
  given in Sect.~\ref{ssec:testarchi}. The training set was kept fixed to the baseline.}
\label{tab:archi}
\end{tiny}
\end{table}

\subsection{Network architectures}
\label{ssec:testarchi}

After characterizing the influence of the ground-truth data set, we used the baseline sets L1 and N1 to compare the performance of various network
architectures. We tested several tens of network architectures obtained from variations of the baseline CNN, G-CNN and ResNet introduced in
Sect.~\ref{sec:archi}, in order to find the best network configurations for classifying small, ground-based lens image cutouts. Below, we highlight
a representative subset of these tests, after excluding all architectures showing poor performance (${\rm AUROC \lesssim 0.9}$).

{\noindent \bf Convolutional neural networks.} The CNNs are adapted from our baseline architecture that was previously applied to lens search in
PanSTARRS multiband images \citep{canameras20}. Apart from the items described below, all network parts are kept fixed to the baseline. To begin,
CNN v2 adds a max-pooling layer after the third convolutional layer to further reduce the dimensions before flattening. CNN v3 includes a fourth
convolutional layer with a kernel size of 5\,$\times$\,5, while adapting the kernels of the first three layers to 11\,$\times$\,11, 9\,$\times$\,9,
and 6\,$\times$\,6. In contrary to v3, CNN v4 removes the third convolutional layer of the baseline architecture. Moreover, CNN v5 discards dropout
regularization, CNN v6 only uses two FC hidden layers with larger number of neurons (1024 each), and CNN v7 uses batch normalization between each
layer. Other architectures tested implement various number of filters in each convolutional layer, various position and number of max-pooling
layers, or different number of neurons in the FC layers. We also tried to suppress all max-pooling layers, to move strides within the convolutional
layers, and to change the kernel sizes in the max-pooling layers. These additional tests either showed minor differences from the baseline CNN or
degraded the performance. Finally, varying dropout rates between 0.1 and 0.7 in steps of 0.1 gave similar results to ${\rm dropout = 0.5}$.  

{\noindent \bf Residual neural network.} The ResNets are variations of the baseline architecture presented in Sect.~\ref{sec:archi}, keeping a
ResNet18-like structure with 8 blocks comprising 2 convolutional layers, batch normalization and ReLU activations. ResNet v2 has a kernel of
5\,$\times$\,5 pixels instead of 3\,$\times$\,3 in the first convolutional layer before the residual blocks. ResNet v3 removes batch normalization
after the first convolutional layer. ResNet v4 lowers the number of feature maps per group of two layers to 16, 32, 64, and 128. With respect to
the baseline, ResNet v5 replaces the 2D average pooling with 2D maximal pooling layer before flattening. Moreover, ResNet v6 uses the lower number
of filters of v4, while also removing the pooling layer and adding a new FC layer of 512 neurons. ResNet v7 sets ${\rm stride = 1}$ instead of 2 in
the second block and adds a FC layer of 128 neurons. Lastly, ResNet v8 tests the effect of applying dropout with rate of 0.5 before each FC layer.

{\noindent \bf Group-equivariant neural networks.} Given the low performance of networks trained with the original G-CNN from \citet{cohen16}, we
used three variants better optimized for the small HSC image cutouts. G-CNN v1 has kernels of 7\,$\times$\,7, 7\,$\times$\,7, 5\,$\times$\,5, and
3\,$\times$\,3 pixels and 10, 10, 20, and 20 features maps per convolutional layer, and 64 neurons in the first FC layer before the single-neuron
output layer. G-CNN v2 increases the number of feature maps to 20, 20, 40, and 40, and the number of neurons in the first FC layer to 128. G-CNN
v3 is same as v1, but with 7\,$\times$\,7 convolutions.

\begin{figure*}
\centering
\includegraphics[width=.49\textwidth]{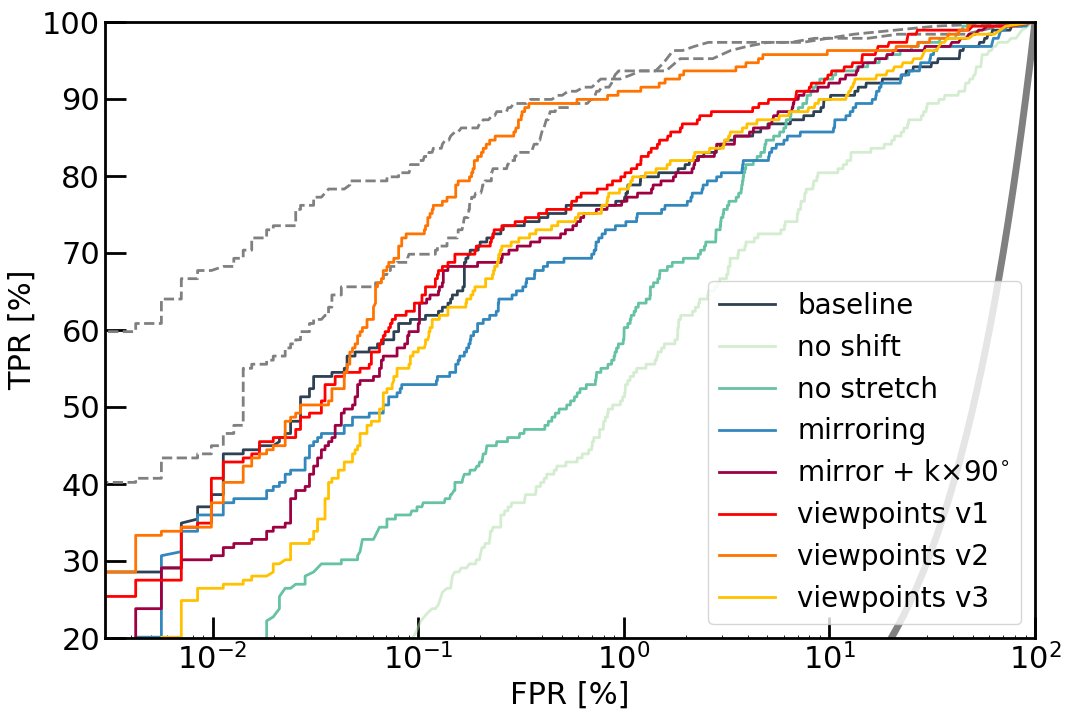}
\includegraphics[width=.49\textwidth]{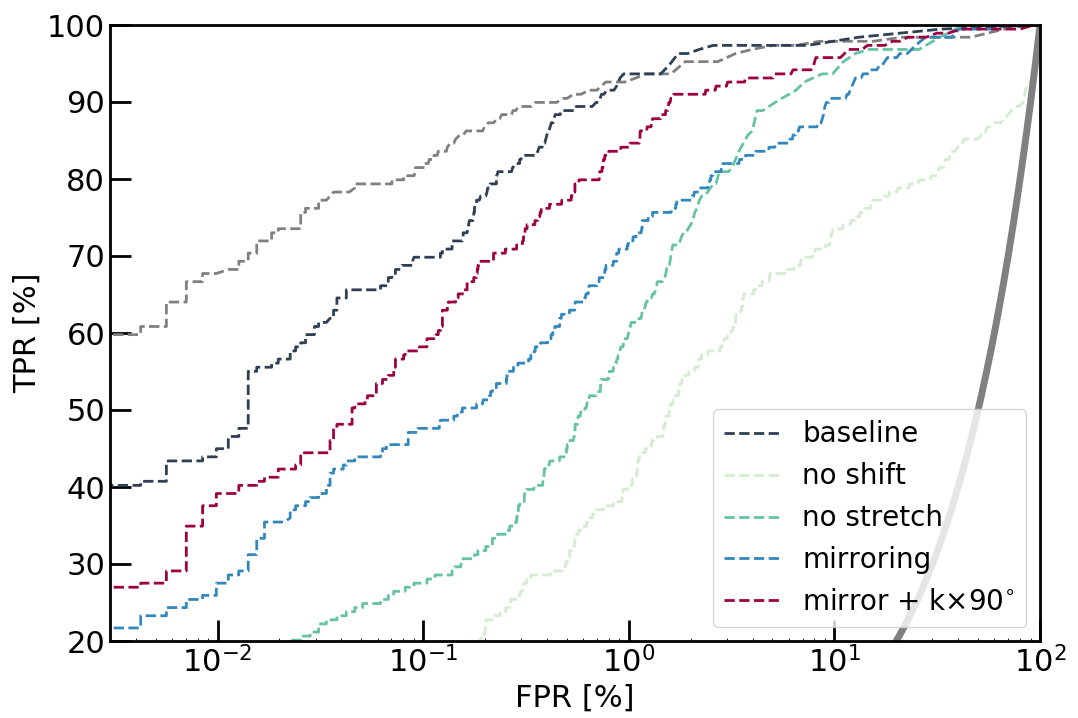}
\caption{Influence of the data augmentation procedure for our baseline data set, and for the baseline CNN (left, solid lines) and ResNet (right, dashed
  lines) architectures. The standard data processing plotted in dark blue consists in applying random shifts to the image centroids and square root
  stretches. The light green curves show networks trained without centroid shifts, and the dark green curves illustrate the performance without square
  root stretch. The blue and brown curves correspond to networks trained on images loaded together with the frames mirrored horizontally and vertically,
  in each of the three $gri$ bands, with (brown) and without (blue) random rotations by $k\times \pi/2$. The red, orange, and yellow curves in the left
  panel show the CNNs trained using viewpoints of the original images as inputs (see Sect.~\ref{ssec:testaug} for details). For reference, the dashed
  grey lines show two good networks (the baseline ResNet and the ResNet from C21). The thick grey curve corresponds to a random classifier.}
\label{fig:aug}
\end{figure*}

\begin{table}
\caption{Tests of data processing and augmentation methods.}
\centering
\begin{tiny}
\begin{tabular}{lcccccc}
\hline
\hline \\[-0.5em]
 & \multicolumn{3}{c}{CNN} & \multicolumn{3}{c}{ResNet} \\
\cmidrule(lr){2-4}\cmidrule(lr){5-7}
Processing & AUROC & TPR$_{\rm 0}$ & TPR$_{\rm 10}$ & AUROC & TPR$_{\rm 0}$ & TPR$_{\rm 10}$ \\[+0.5em]
\hline \\[-0.5em]
baseline & 0.9557 & 0.0 & 44.4 & {\bf 0.9913} & {\bf 36.0} & {\bf 55.0} \\
\hline \\[-0.5em]
no shift & 0.9071 & 0.0 & 0.0 & 0.8332 & 0.0 & 0.0 \\
no stretch & 0.9660 & 0.0 & 18.5 & 0.9746 & 7.4 & 16.9 \\
mirror & 0.9505 & 5.3 & 38.1 & 0.9699 & 16.9 & 31.2 \\
mirror + $k \pi /2$ & 0.9686 & 9.0 & 32.3 & 0.9834 & 22.2 & 40.2 \\
viewpoints v1 & 0.9773 & 17.5 & 43.4 & \dots & \dots & \dots \\
viewpoints v2 & 0.9841 & 11.1 & 42.3 & \dots & \dots & \dots \\
viewpoints v3 & 0.9632 & 5.3 & 27.5 & \dots & \dots & \dots \\
\hline
\end{tabular}
\tablefoot{Each test is described in Sect.~\ref{ssec:testaug}. The baseline data set is used together with the baseline CNN and ResNet. The
  extraction of viewpoints was not tested for the baseline ResNet because this architecture is poorly suited for small input frames.}
\label{tab:aug}
\end{tiny}
\end{table}

The performance with these various architectures are summarized in Fig.~\ref{fig:archi} and Table~\ref{tab:archi}. Overall, we obtain the best
AUROC, TPR$_{\rm 0}$, and TPR$_{\rm 10}$ for the baseline ResNet. Such networks reaching the highest TPR$_{\rm 0} \simeq 30$--40\% are most useful to
real lens searches in strongly unbalanced data sets, as they allow to drastically limit the number of contaminants and to save significant human
inspection time. While only the baseline ResNet reach such elevated recall at zero contamination, all three G-CNNs and some of our CNNs and ResNets
reach high TPR$_{\rm 10}$ of about 50\%.

In the literature, while ResNets have been often used for lens finding \citep[e.g.,][]{huang20,shu22}, their performance have been essentially
evaluated on the Euclid lens-finding dataset \citep{metcalf19}. The ResNet from \citet{lanusse18} won the ground-based part of the challenge but
\citet{schaefer18} found that deeper networks do not necessarily provide better performance. We can extend this comparison based on our new test
sets drawn from real data and including representative populations of contaminants. Overall the best performance are obtained with ResNets but the
improvement of ResNet architectures with respect to CNNs in terms of AUROC is clearly not systematic. Only the values of TPR$_{\rm 0}$ tend to be
higher for the ResNet ($\simeq 10$--40\%), with none of the CNN architecture we tested exceeding TPR$_{\rm 0} \simeq 10$\%.

In particular, we find that CNNs with additional layers do not improve the classification. Using only two convolutional layers apparently degrades
the performance (CNN v4), but we also noticed that changing the kernel size and lowering the number of filters allowed us to recover metrics similar
to the baseline CNN. Moreover, our results show that fine-tuning the ResNet architecture helps improve the performance with respect to the original
ResNet18. Optimizing the structure of the first layer in these ResNets to the size of input images appears to be particularly important given the
lower AUROC and TPR$_{\rm 0}$ in ResNets v2 and v3 with respect to the baseline. Similarly, varying the last FC layers has substantial impact (ResNets
v7 and v8). The decrease in AUROC for ResNet v4 nonetheless shows that the number feature maps in the original ResNet18 architecture is well-suited
for our classification problem. Finally, the three G-CNNs give remarkably stable performance in Table~\ref{tab:archi}, as well as for alternative
training sets, in agreement with the lower generalization gap observed in their loss curves compared to CNNs and ResNets.

\subsection{Data processing}
\label{ssec:testaug}

The way a given ground truth data set is loaded can affect the properties of the feature representation learned by the network, and thus impact
its classification performance. Given the possibility to obtain realistic simulations, our lens-finding problem differs from other classification
tasks working in low data regimes \citep[e.g., radio galaxy classification,][]{aniyan17,slijepcevic22}. For lens searches, data augmentation is
mainly intended to help models make stable predictions despite the presence of small perturbations. Here, we tested the impact of various data
augmentation schemes using the baseline data set and the three $gri$ bands. The standard data processing recipe for both the baseline CNN and the
baseline ResNet consists in applying random shifts sampled uniformly between $-$5 and +5 pixels to the image centroids, and square root stretches
after clipping pixels with negative values to zero. We demonstrate the importance of these two transformations in optimizing the performance, and
we test a non-exhaustive list of additional data augmentation techniques.

\begin{figure}
\centering
\includegraphics[width=.49\textwidth]{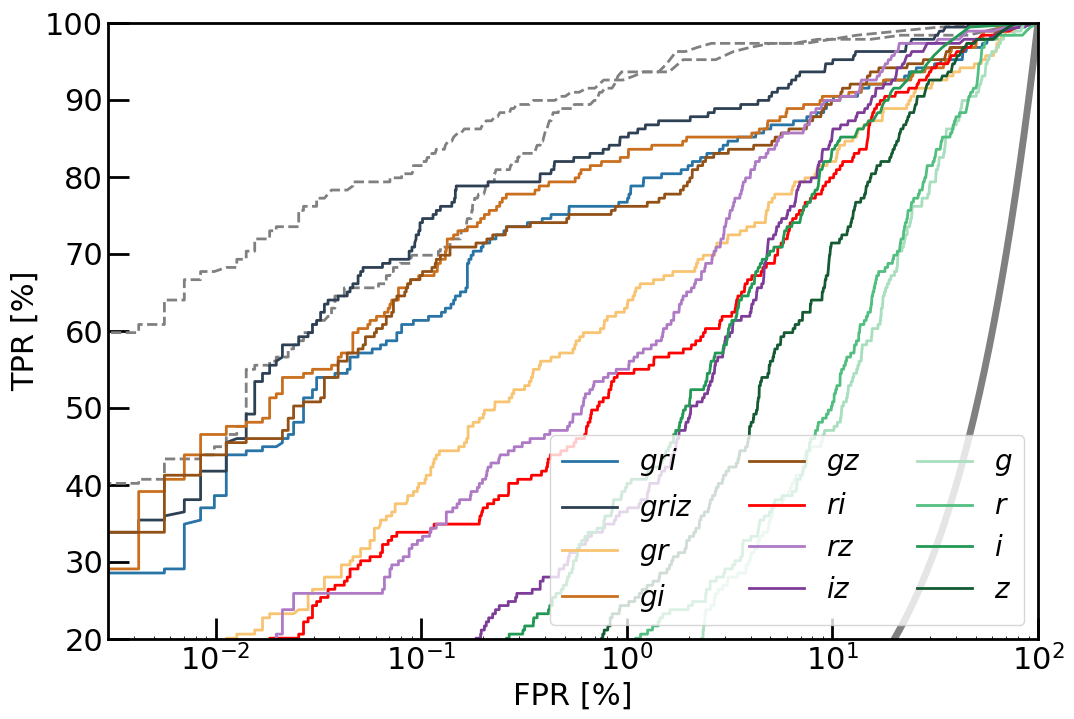}
\caption{ROC curves for training with different numbers of observing bands, for the baseline CNN and training set. For reference, the dashed grey
  lines show two good networks (the baseline ResNet and the ResNet from C21). The thick grey curve corresponds to a random classifier. Adding
  $z$-band to the standard $gri$ three-band input helps increase the AUROC.}
\label{fig:numbands}
\end{figure}

{\noindent \bf Centroid shifts.} The networks were trained and tested on images perfectly centered on the relevant galaxy. The ROC curves in
Fig.~\ref{fig:aug} show a significant drop in performance when removing the random shift in the image centroid. The AUROC of 0.9913 obtained for
the baseline ResNet decreases to 0.8332 for this scenario, and both CNN and ResNet trained without random shifts have TPR$_{\rm 0}$ of 0\%. These
results suggest that, in these cases, neural networks learn spatial offsets as a determinant feature for galaxy classification. 

{\noindent \bf Stretching and normalization.} To show the importance of applying a square root stretch to individual images, we trained and tested
the networks on original images from the baseline data set. The lower AUROC, TPR$_{\rm 0}$, and TPR$_{\rm 10}$ obtained for these networks with respect
to the baseline (see Table~\ref{tab:aug}) demonstrates the benefit of stretching the pixel scale to boost the low-luminosity features and help learn
the relevant information encoded in the lensed arcs. We tested other networks trained on alternative data sets from Sect.~\ref{ssec:train} and found
systematically better performance with square root stretch. On the contrary, using a log-scale or other approaches did not provide any improvement.
A range of image normalization techniques were also tested such as scaling pixel values to the range 0--1, or normalizing images to zero mean and
unit variance, and these techniques were applied either to individual images or to image batches. These processing methods were discarded from
Fig.~\ref{fig:aug} as none of them reached performance comparable to the baseline networks.

{\noindent \bf Rotation and mirroring.} First, we augmented the data set by loading the frames mirrored horizontally and vertically together with
the original images, which resulted in three input images per object per band. Second, we applied random $k \times 90^{\circ}$\ rotations to each
input object, while also loading the original and mirrored frames as described above. In both cases, the objective is to help the networks learn
invariance with respect to rotation and flipping operations. For the CNN, these approaches result in AUROC and TPR$_{\rm 10}$ comparable to the
baseline with a minor increase in TPR$_{\rm 0}$ up to $\simeq 5$--10\%. For the ResNet, performance are slightly lower than the baseline.

{\noindent \bf Image viewpoints.} Inspired by the methodology employed in \citet{dieleman15} to reduce overfitting and improve rotation invariance
for their galaxy morphological classifier, we tested using viewpoints (image crops) of the original images as input. In the first version, we used
four viewpoints of 40\,$\times$\,40 pixels and random centers, corresponding to cropped version of the original images of 60\,$\times$\,60 pixels in
the baseline data set. Centroid positions are kept identical between bands. Depending on their position, these viewpoints cover either a fraction or
the totality of the relevant lens systems and nonlens galaxies, and they have a significant mutual overlap. In that case, for a given entry in the
data set, the neural networks are fed with a total of 12 input frames, corresponding to four viewpoints per band. In the second version, eight
randomly-centered viewpoints of 40\,$\times$\,40 pixels were loaded, and we applied a random rotation by $k \times 90^{\circ}$ to each viewpoint. In
the third version, we also used eight viewpoints with random centroids and random rotations, but increased their size to 52\,$\times$\,52 pixels.
Given that our ResNet architectures are not adapted to these smaller crops, we only tested the extraction of viewpoints with the baseline CNN.
Fig.~\ref{fig:aug} and Table~\ref{tab:aug} show that these approaches help boost the AUROCs, and interestingly, the TPR$_{\rm 0}$ values are
systematically higher than the baseline, and up to 17.5\% for the first version.

\begin{table}
\caption{Performance as a function of the number of observing bands.}
\centering
\begin{tiny}
\begin{tabular}{lccc}
\hline
\hline \\[-0.3em]
Bands & AUROC & TPR$_{\rm 0}$ & TPR$_{\rm 10}$ \\[+0.5em]
\hline \\[-0.5em]
baseline ($gri$ bands) & 0.9557 & 0.0 & 44.4 \\
\hline \\[-0.5em]
$griz$ & {\bf 0.9820} & 13.8 & {\bf 49.2} \\
$gr$ & 0.9297 & 0.0 & 20.6 \\
$gi$ & 0.9604 & 24.3 & 48.2 \\
$gz$ & 0.9601 & {\bf 30.7} & 46.0 \\
$ri$ & 0.9322 & 8.5 & 16.4 \\
$rz$ & 0.9606 & 0.0 & 16.4 \\
$iz$ & 0.9401 & 0.5 & 4.8 \\
$g$ & 0.8272 & 0.0 & 0.5 \\
$r$ & 0.8382 & 0.0 & 1.1 \\
$i$ & 0.9392 & 0.0 & 5.3 \\
$z$ & 0.9022 & 0.0 & 0.5 \\
\hline
\end{tabular}
\tablefoot{These tests all use the baseline CNN and baseline training set.}
\end{tiny}
\label{tab:numbands}
\end{table}

\subsection{Number of observing bands}

The influence of the observing bands was tested using the three $gri$ bands as baseline, and comparing with predictions obtained for combinations
of one, two, or four bands. In HSC Wide, $gri$ bands have the best 5$\sigma$ point-source sensitivities of 26.6, 26.2, and 26.2 mag, respectively
\citep{aihara19}, together with a remarkably consistent depth between bands. The $gri$ bands will also be the deepest in the 10-year LSST stacks,
about 1~mag deeper than HSC Wide according to the LSST baseline design \citep[][]{ivezic19}. The $z$-band also considered in our analysis has a
depth of 25.3~mag, and it will also be $\simeq$1~mag shallower than $gri$ in the final LSST stacks. For strong lenses, this redder band can play
a role in identifying the signatures from the foreground galaxies with limited contamination from background lensed arcs which are mostly blue.
In the LSST era, additional $u$-band images will reach similar depth as in the $z$-band, and they will play a role in identifying strong lenses.
The results of our tests in Fig.~\ref{fig:numbands} and Table~\ref{tab:numbands} show that training the CNN jointly with the four bands helps boost
the AUROC, TPR$_{\rm 0}$ and TPR$_{\rm 10}$ compared to the baseline CNN trained on $gri$. Interestingly, training either with $gi$ or $gz$ bands
gives better performance than $gri$, with the highest TPR$_{\rm 0}$ of 30.7\% obtained with $gz$ bands. This could either be due to different image
resolutions per band (see Sect.~\ref{sec:discu}), or to fluctuations in TPR$_{\rm 0}$ due to the moderate number of test lenses. Other combinations
of two bands give either poor ${\rm AUROC \simeq 0.93}$ or very low recall at zero contamination. Lastly, training with one band performs much
worse, with the highest AUROC obtained for the $i$-band, and none of the single-band networks providing high enough TPR$_{\rm 10} \geq 10$\%.

\begin{table}
\caption{Performance for various fractions of the overall data set.}
\centering
\begin{tiny}
\begin{tabular}{lcccccc}
\hline
\hline \\[-0.5em]
 & \multicolumn{3}{c}{CNN} & \multicolumn{3}{c}{ResNet} \\
\cmidrule(lr){2-4}\cmidrule(lr){5-7}
Fraction & AUROC & TPR$_{\rm 0}$ & TPR$_{\rm 10}$ & AUROC & TPR$_{\rm 0}$ & TPR$_{\rm 10}$ \\[+0.5em]
\hline \\[-0.5em]
10\% & 0.9806 & 0.0 & 41.3 & 0.9614 & 5.8 & 38.1 \\
20\% & 0.9582 & 0.0 & 51.9 & 0.9532 & 3.7 & 16.9 \\
30\% & 0.9467 & 0.0 & 28.0 & 0.9675 & 12.7 & 36.0 \\
40\% & 0.9426 & 0.0 & 33.9 & 0.9063 & 16.4 & 27.0 \\
50\% & 0.9530 & 0.0 & 38.6 & 0.8583 & 6.4 & 22.2 \\
60\% & 0.9661 & 0.0 & 42.3 & 0.9568 & 15.3 & 24.3 \\
70\% & 0.9623 & 0.0 & 45.5 & 0.9637 & 21.2 & 37.6 \\
80\% & 0.9592 & 0.0 & 44.9 & 0.9686 & 25.9 & 45.0 \\
90\% & 0.9483 & 5.8 & 31.8 & 0.9820 & 8.5 &  44.4 \\
100\% & 0.9557 & 0.0 & 44.4 & {\bf 0.9913} & {\bf 36.0} & {\bf 55.0} \\
\hline
\end{tabular}
\tablefoot{The baseline training set is used together with the baseline CNN and ResNet.}
\label{tab:datasize}
\end{tiny}
\end{table}

\subsection{Data set size}

The ideal size of the ground-truth data set depend both on the network depth and size of input images. While classical CNNs with only a couple of
convolutional layers show stable performances for training sets with $\geq 10^5$ images \citep[e.g.,][]{he20}, the classification accuracies for
smaller sets down to $\simeq 10^4$ examples deserve further tests, in particular for deeper ResNet architectures. We characterized the influence
of the data set size by training and validating our CNN and ResNet on different fractions of the overall baseline data set. We used between 10\%
and 100\% of all images, in steps of 10\%, or between 7,000 and 70,000 training examples. Table~\ref{tab:datasize} (and Fig.~\ref{fig:datasize})
show that AUROCs do not smoothly increase as a function of the data set size. The results are more stable for CNNs which have AUROC in the range
0.9426--0.9806, TPR$_{\rm 0} \simeq 0$\%, and TPR$_{\rm 10} \simeq 30$--50\%. The CNN with highest AUROC uses only 10\% of the data set, suggesting
that $\simeq 10^4$ examples is sufficient to train shallow networks. Many ResNets have better performance than the CNNs, with AUROC up to 0.9913
and zero-contamination recall up to 36.0\% for the ResNet using the entire data set, but the scatter is also larger (e.g. AUROC of 0.8583 for a
fraction of 50\%). In contrast to CNNs, the ResNets tend to show an improvement in performance as a function of the number of training examples
(see Table.~\ref{tab:datasize}). This trend is most prominent for the TPR$_{\rm 0}$ values which increase from 3.7--16.4\% for fractions $\leq 50$\%
to 8.5--36.0\% for fractions $> 50$\%.

\subsection{Difference images}
\label{ssec:testdiffima}
  
\begin{figure}
\centering
\includegraphics[width=.49\textwidth]{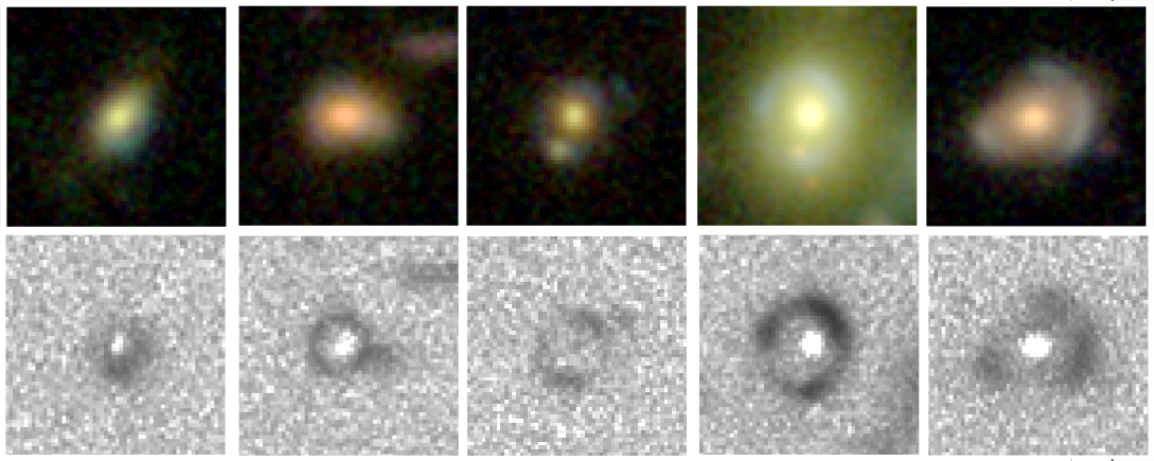}
\caption{Mosaic of difference images for mocks in the baseline data set. {\it Top:} Three-color images from the original $gri$ stacks.
  {\it Bottom:} The corresponding difference images based on a simple subtraction of the rescaled $i$-band to the $g$-band frames.}
\label{fig:diffima_mosaic}
\end{figure}

The populations of lens and source galaxies targeted by our overall search experiment show a strong color dichotomy. As we specifically focus
on foreground galaxies with the highest lensing cross-section, samples of lens galaxies are dominated by massive early-type galaxies with red
colours and smooth light profiles. For deep, ground-based imaging surveys, typical galaxies magnified by these light deflectors are located at
$z \simeq 1$--4, an epoch dominated by bluer star-forming galaxies. For galaxy-scale systems, the signals from both components are necessarily
blended to some extents. To help the networks access the signal from the lensed arcs, we attempted using difference images (DI) obtained from
the subtraction of the red $i$-band to the blue $g$-band frames. The subtraction $g - \alpha i$ includes a rescaling term $\alpha$ computed as
follows:
\begin{equation}
\log \alpha = 0.4 \times (i_{\rm cModel}-g_{\rm cModel})
\end{equation}
where cModel are the composite model magnitudes obtained from the combined fit of exponential and de Vaucouleurs profiles \citep{lupton01,bosch18}.
Because it is based on the cModel photometry, $\alpha$ accounts for the PSF FWHM in each band. To compute the difference image of each object in
the baseline data set and in the test set, we used the cModel magnitudes listed in the HSC PDR2 tables. This approach only considers the central
object, and difference images of the mocks were thus obtained using the photometry of the lens galaxy without arc-light contamination. For mocks
in the baseline data set, Fig.~\ref{fig:diffima_mosaic} shows that difference images efficiently remove the lens light and help emphasize the
signatures from background sources. The PSF FWHM of $g$ and $i$-band images were not matched to avoid convolution overheads and, in consequence,
the central emission is oversubtracted. This artefact is mitigated by the clipping of negative pixels included in our data loader.

Figure~\ref{fig:diffima} and Table~\ref{tab:diffima} show the results of training the networks on the difference images alone, or after combining
these frames with original images in $i$ band, which has the best seeing, or with the three $gri$ bands. We find poor classifications for networks
trained, validated, and tested only with difference images, either due to the lack of signal from the lens galaxy and companions, to residuals from
differences in $g$ and $i$-band seeing, or to both effects. Other approaches depend on the architecture. For the CNN, we obtain higher AUROC,
TPR$_{\rm 0}$, and TPR$_{\rm 10}$ when using DI $+$ $gri$ bands compared to the baseline. This suggests that adding input frames with clearer signal
from the lensed arcs help slightly improve the recall of SuGOHI lenses. However, this gain disappears for the ResNet.

\subsection{Masking neighbors}
\label{ssec:testmask}

Our classification experiments include neighboring galaxies in the 10\arcsec\,$\times$\,10\arcsec\ HSC cutouts around the central galaxy or
strong-lens system tabulated in our data sets. Neural networks learn the status of these nearby, unassociated galaxies during the training phase.
We tried to mask the neighboring sources as part of the pre-processing of images in the training, validation, and test sets, to determine whether
or not this helps the networks focus on the relevant sources and improve their performance. We used {\tt SExtractor} \citep{bertin96} to mask
galaxies well deblended from the central objects, which were required to be within 5 pixels from the cutout centers. A low number of deblending
thresholds, {\tt DEBLEND\_NTHRESH} of 16 and a contrast parameter, {\tt DEBLEND\_MINCONT} of 0.01 were chosen to avoid identifying local peaks in
the light distributions. We defined the masks in $r$ band, using the 3$\sigma$ isophotes after convolving the images by Gaussian kernels with FWHM
of 2 pixels. This optimal tradeoff provided adequate deblending in all three $gri$ bands, while smoothing the mask edges, and keeping features near
the central galaxy (e.g. lensed arcs) within the masks. In some cases, interesting features are inevitably masked out with this automated procedure.
We nonetheless noticed that all multiple images are within the masks for nearly all SuGOHI test lenses, and all strong-lens simulations with
$\theta_{\rm E} \lesssim 2.0$\arcsec. The lensed arcs with largest separation from the lens center are masked out for only $\simeq 10$\% of mocks
with $\theta_{\rm E} > 2.0$\arcsec.

\begin{figure}
\centering
\includegraphics[width=.49\textwidth]{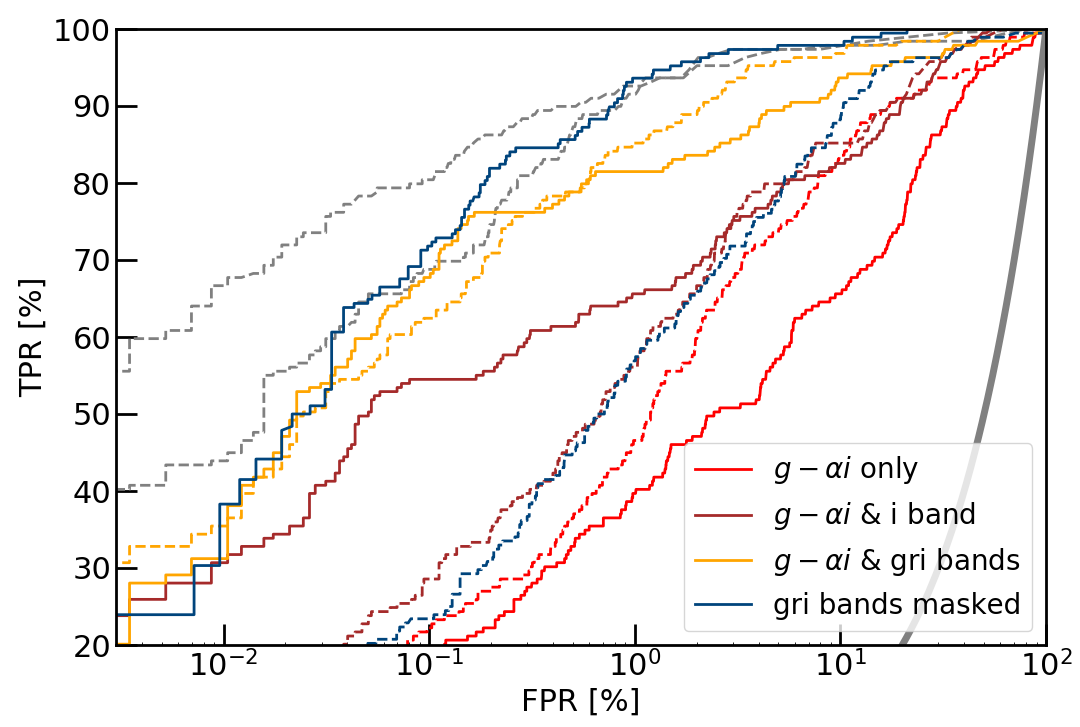}
\caption{Performance of networks trained, validated and tested on difference images $g - \alpha i$ alone (red), joined with images from the $i$
  band that has best seeing (brown), or joined with the three $gri$ bands (orange). Blue curves show the baseline CNN (solid) and ResNet (dashed)
  trained, validated and tested on $gri$ images with neighboring galaxies masked (see Sect.~\ref{ssec:testmask}). For reference, the dashed grey
  lines show two good networks (the baseline ResNet and the ResNet from C21). The thick grey curve corresponds to a random classifier. We obtain
  a significant improvement in AUROC for the CNN after masking companions galaxies.}
\label{fig:diffima}
\end{figure}

The results in Fig.~\ref{fig:diffima} and Table~\ref{tab:diffima} indicate that the masking procedure improves the metrics for the CNN, but not for
the ResNet. For the CNN, we obtain a substantial increase in AUROC to 0.9949, which is the highest AUROC over all tests conducted in this study, and
we get a TPR$_{\rm 10}$ of 50\% approaching the recall of the baseline ResNet. For the ResNet, the lower performance compared to the baseline might
suggest that artefacts from the masking procedure (e.g., sharp mask edges, truncated galaxy light profiles) inevitably affect the underlying model.

\subsection{Number of output classes}

To assess the influence of the number of ouput classes in the neural networks used for our classifications, we tested the baseline CNN and ResNet
architectures with four classes instead of two. The data set was kept similar to the baseline, with the first class including the baseline mocks
from set L1, and the three other classes containing LRGs, spirals, and random galaxies, respectively, selected in the same way as for set N1. Given
the moderate number of compact groups in our parent sample, this type of nonlens galaxies was discarded from this test. The number of elements per
class was kept balanced to 24,500 examples, and architectures matched the baseline CNN and ResNet apart from the four-neuron output layer. We
determined the performance of this multiclass classification using the ROC curves of class ``lens''. For the CNN, we obtain AUROC, TPR$_{\rm 0}$,
and TPR$_{\rm 10}$ of 0.9808, 12.2\%, and 33.3\%, respectively, while for the ResNet we find an AUROC of 0.9824, a TPR$_{\rm 0}$ of 9.0\%, and a
TPR$_{\rm 10}$ of 31.8\%. Compared to binary classification, the CNN reaches higher AUROC and TPR$_{\rm 0}$ but lower TPR$_{\rm 10}$, and the performance
globally decreases for the ResNet.

\begin{table}
\caption{Performance of networks using difference images.}
\centering
\begin{tiny}
\begin{tabular}{lcccccc}
\hline
\hline \\[-0.5em]
 & \multicolumn{3}{c}{CNN} & \multicolumn{3}{c}{ResNet} \\
\cmidrule(lr){2-4}\cmidrule(lr){5-7}
Type & AUROC & TPR$_{\rm 0}$ & TPR$_{\rm 10}$ & AUROC & TPR$_{\rm 0}$ & TPR$_{\rm 10}$ \\[+0.5em]
\hline \\[-0.5em]
baseline $gri$ & 0.9557 & 0.0 & 44.4 & 0.9913 & {\bf 36.0} & {\bf 55.0} \\
\hline \\[-0.5em]
DI only & 0.8780 & 0.5 & 4.2 & 0.9322 & 1.1 & 11.6 \\
DI + $i$ band & 0.9495 & 7.9 & 34.4 & 0.9530 & 3.7 & 14.3 \\
DI + $gri$ & 0.9684 & 9.5 & 45.0 & 0.9883 & 24.3 & 42.9 \\
\hline \\[-0.5em]
$gri$ + masks & {\bf 0.9949} & 0.0 & 50.0 & 0.9559 & 7.45 & 14.9 \\
\hline
\end{tabular}
\tablefoot{The middle three rows list networks trained on difference images from the baseline data set, with metrics evaluated on the difference
  images from the test set. The last networks use the original $gri$ stacks, after masking surrounding galaxies within the cutouts. Further details
  are given in Sect.~\ref{ssec:testdiffima} and \ref{ssec:testmask}.}
\label{tab:diffima}
\end{tiny}
\end{table}

\subsection{Committees of networks}
\label{ssec:ens}

Ensemble learning refers to training multiple neural networks for the same task and combining predictions from the ensemble of models. This
method has proven helpful to mitigate the stochasticity of the learning process, and to lower the generalization error and variance in output
scores \citep{hansen90,krogh94}. We applied this approach to our binary classification problem, first by training a committee of five networks
with fixed architecture, ground truth dataset and dataset split, but with different weight initialisation \citep[e.g.,][]{schaefer18}. Secondly,
we trained five networks with fixed architecture and ground truth data, but with random split into training and validation, and with different
random initialization of the networks weights. In both cases, the procedure was repeated for the CNN and ResNet. Thirdly, we used committees of
the best networks from Table~\ref{tab:dataset} trained on different ground-truth data sets (see Sect.~\ref{ssec:testdata}). The output
scores of the network committees were then averaged to obtain the final prediction. Other combinations of scores, for instance by taking the
median, minimal, or maximal value over the five models, did not improve the performance.

\begin{figure}
\centering
\includegraphics[width=.49\textwidth]{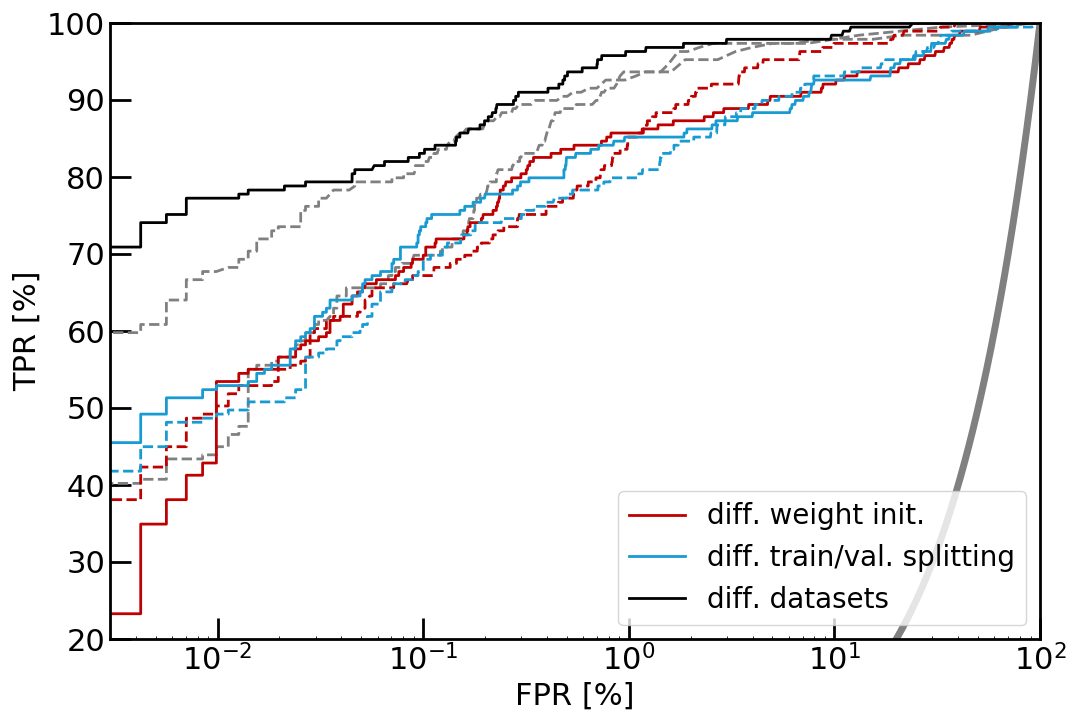}
\caption{Performance of committees of networks. The red curves show the committee of five networks with different weight initialisation. The blue
  curves instead use fixed architecture and ground-truth data, but random split into training and validation, and different weight initialization.
  Black curves show a committee of networks trained on different data sets (see Sect.~\ref{ssec:ens}). As in previous plots, solid and dashed lines
  correspond to the CNN and ResNet architectures, respectively. The third approach gives the most significant boost in performance, and allows to
  overcome the AUROC, TPR$_{\rm 0}$ and TPR$_{\rm 10}$ of the best individual networks shown as grey lines (the baseline ResNet and the ResNet from C21).}
\label{fig:ens}
\end{figure}

Figure~\ref{fig:ens} and Table~\ref{tab:ens} show that the third approach gives the most significant boost in performance. While averaging scores
from random networks in Table~\ref{tab:dataset} does not systematically improve the overall metrics, selecting the networks with good performance
and different internal representations turns out to be more useful. To identify the best networks to combine, we selected the 10 networks with
highest AUROC in Table~\ref{tab:dataset}, and we counted the number of false positives and false negatives overlapping between pairs of networks.
For that, we first selected for each network the 50 SuGOHI lenses with lowest scores and found that the overlap between pairs of networks varied
between 30 and 40 objects. We then identified for each network the 50 nonlenses in COSMOS with highest scores and found that networks tend to be
contaminated by different false positives, with only about 3--15 objects in common between pairs of networks. Given the higher overlap between false
negatives found by each network than between false positives, we averaged the scores from the networks with the lowest number of false positives in
common. We obtained the most significant improvement by combining the baseline ResNet, the ResNet trained on sets L4 and N1, and the ResNet from
\citetalias{canameras21} trained on L7 and N1 (see Fig.~\ref{fig:ens}). This committee reaches the highest TPR$_{\rm 0}$ and TPR$_{\rm 10}$ over our
various tests, and obtaining a recall at zero contamination as high as $\simeq 60$\% opens promising perspectives for pure selection of strong
lenses without human input. Regarding the first two approaches, Table~\ref{tab:ens} shows a significant improvement of each metric for the CNN
\citep[as also found by][]{schaefer18}, but lower performance for the ResNet.

\begin{table}
  \caption{Performance of committees of networks.}
\centering
\begin{tiny}
\begin{tabular}{lcccccc}
\hline
\hline \\[-0.5em]
 & \multicolumn{3}{c}{CNN} & \multicolumn{3}{c}{ResNet} \\
\cmidrule(lr){2-4}\cmidrule(lr){5-7}
 Variations & AUROC & TPR$_{\rm 0}$ & TPR$_{\rm 10}$ & AUROC & TPR$_{\rm 0}$ & TPR$_{\rm 10}$ \\[+0.5em]
\hline \\[-0.5em]
baseline & 0.9557 & 0.0 & 44.4 & 0.9913 & 36.0 & 55.0 \\
\hline \\[-0.5em]
diff. init. & 0.9719 & 16.4 & 55.0 & 0.9881 & 15.9 & 52.9 \\
diff. splitting & 0.9717 & 27.5 & 53.4 & 0.9725 & 24.9 & 50.8 \\
diff. datasets & \dots & \dots & \dots & {\bf 0.9962} & {\bf 59.3} & {\bf 78.3} \\
\hline
\end{tabular}
\tablefoot{Committees of networks with averaging of output scores. The committee of networks trained of different data sets is only shown for the
  best combination obtained with the ResNet architecture (see Sect.~\ref{ssec:ens}).}
\label{tab:ens}
\end{tiny}
\end{table}

\section{Discussion}
\label{sec:discu}

After comparing the performance of various networks using fixed test sets, we explore the ingredients to reduce the dependence on image quality
and orientation.

\subsection{Dependence on local seeing and depth}

In the Wide HSC layer, the excellent $i$-band seeing required for weak-lensing analyses results in significant differences with other bands.
Training our supervised neural networks on these multiband images without matching the PSF per band can introduce dependencies on systematic
variations in seeing FWHM \citep[see also][]{petrillo19a,li20}. In our experiments, all training images are distributed randomly over the PDR2
footprint and, for the lens simulations, local PSF models are used to paint lensed arcs with realistic angular resolutions. Although this procedure
guarantees that the network inputs account for variations in image quality, it does not rule out biases in the output scores as a function of local
seeing FWHMs.

\begin{figure*}
\centering
\includegraphics[height=.25\textwidth]{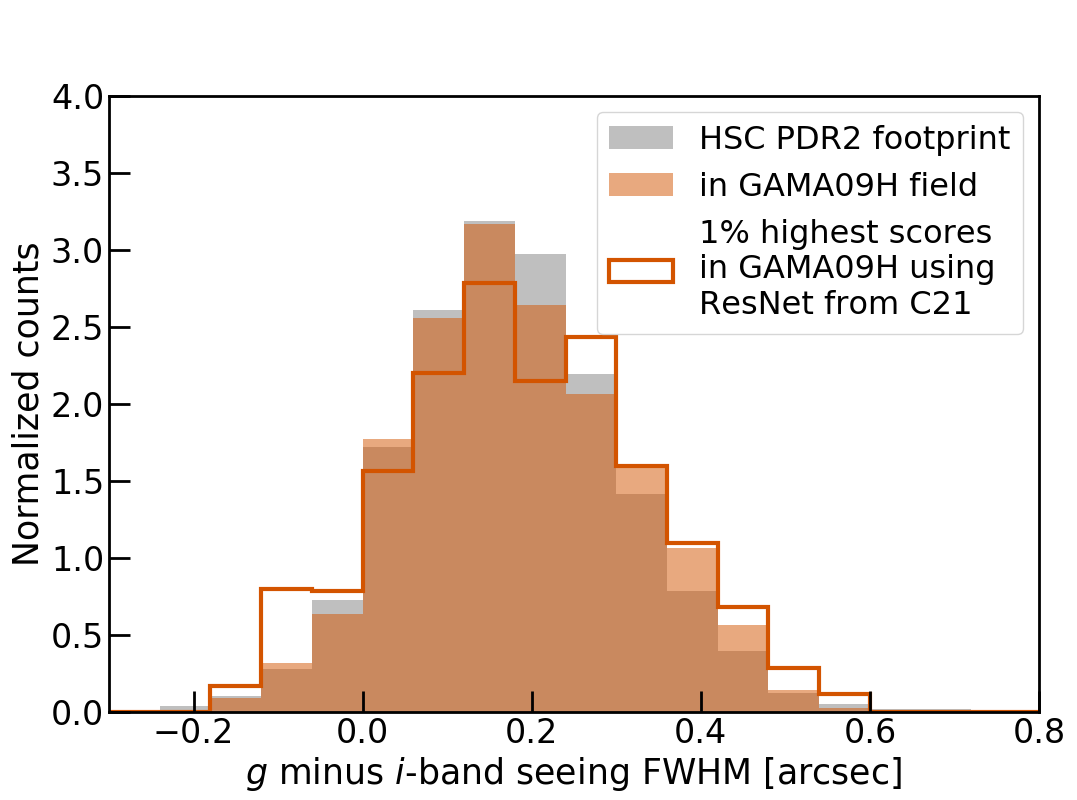}
\includegraphics[height=.25\textwidth]{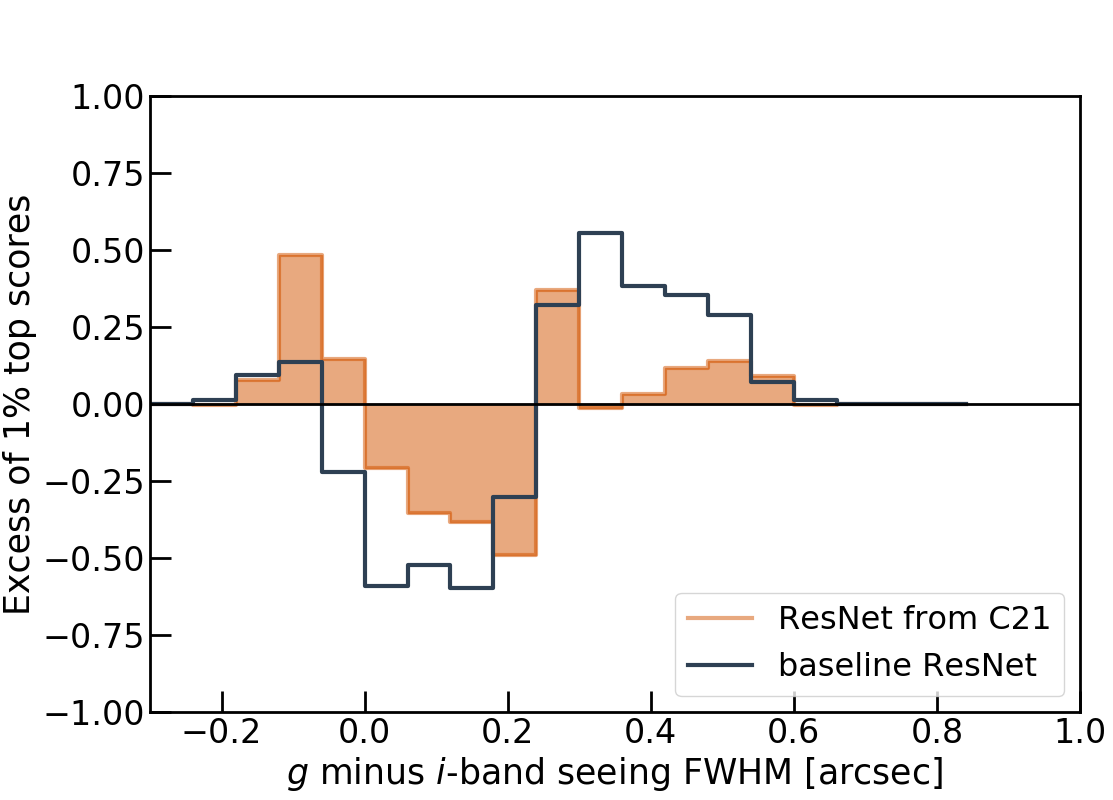}
\includegraphics[height=.25\textwidth]{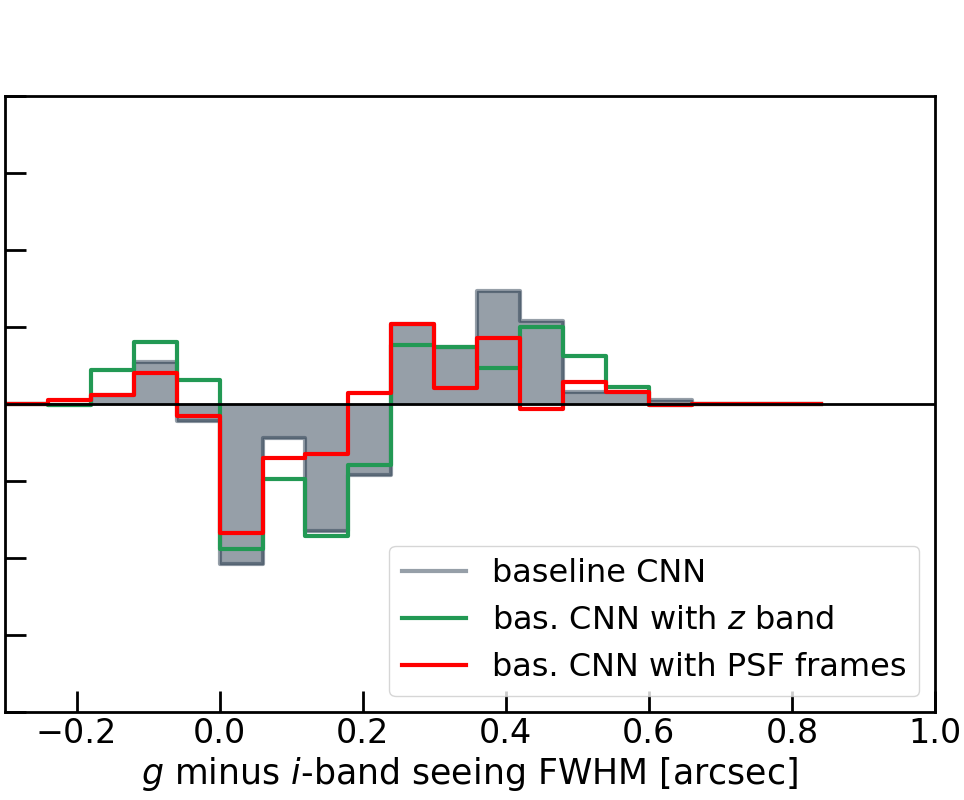}

\includegraphics[height=.25\textwidth]{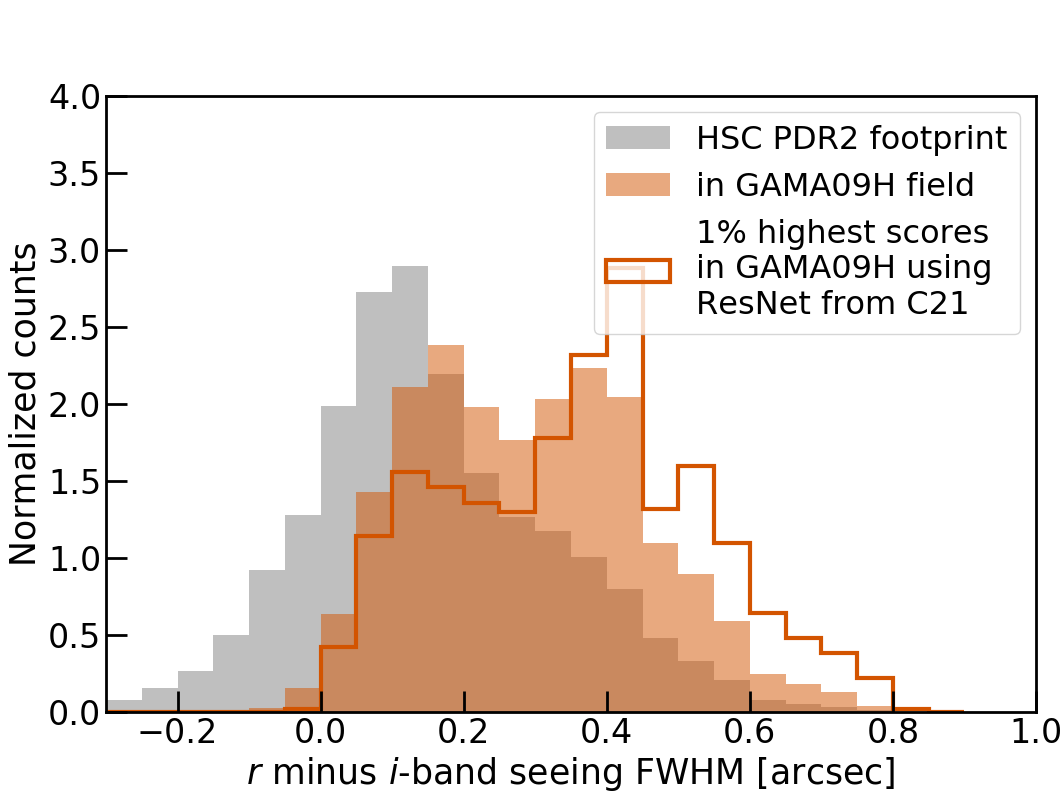}
\includegraphics[height=.25\textwidth]{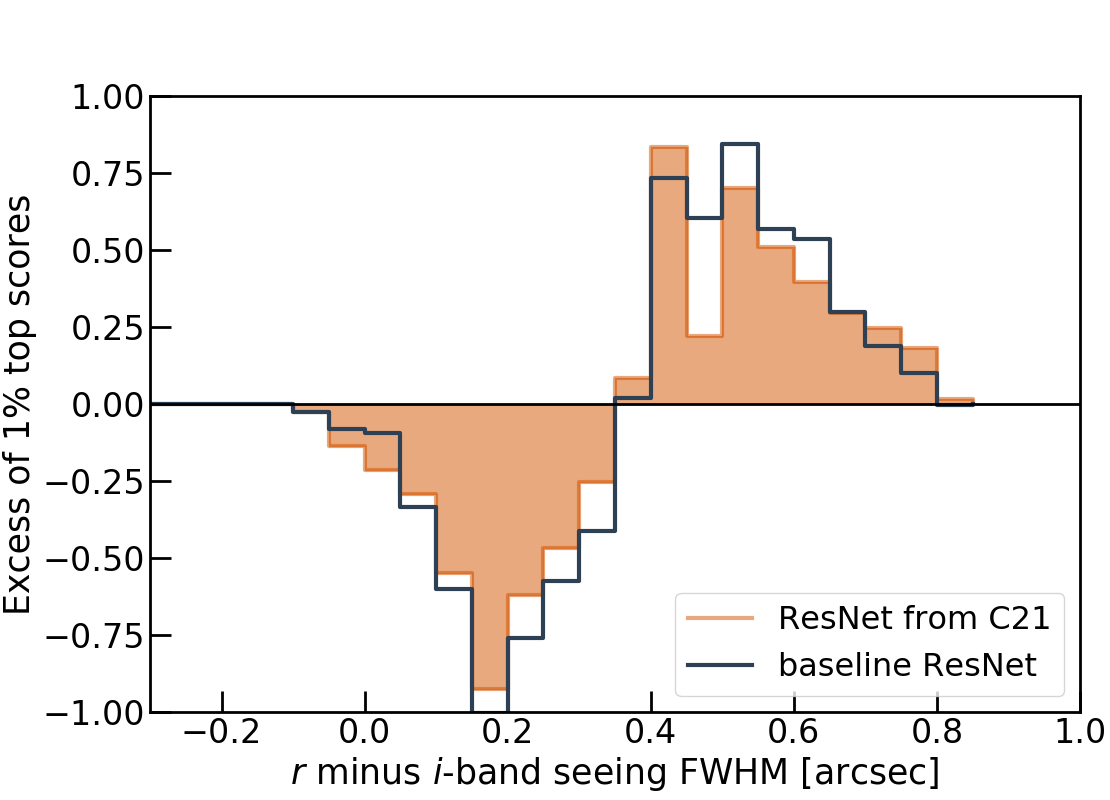}
\includegraphics[height=.25\textwidth]{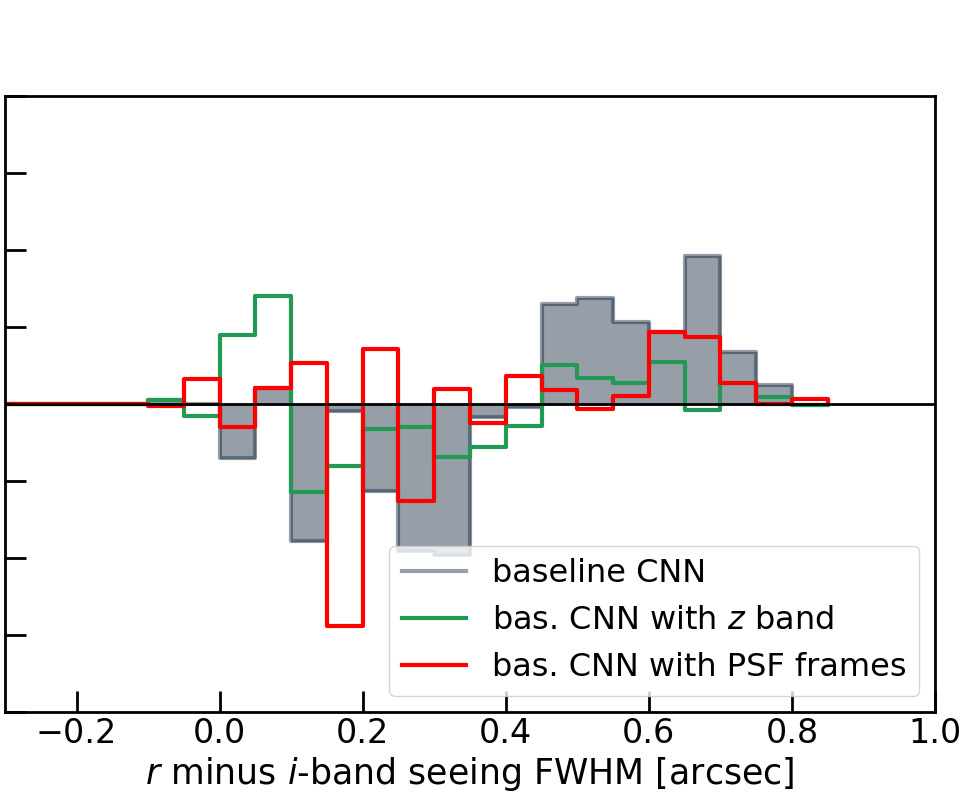}
\caption{Histograms of seeing FWHM difference between pairs of bands. In the left panels, grey histograms show the average distributions over the
  entire HSC PDR2 footprint, and orange filled histograms show the distributions restricted to the GAMA09H field. In addition, step histograms show
  the distributions for galaxies within GAMA09H assigned the 1\% highest scores by the ResNet from \citetalias{canameras21}. In the middle panels,
  orange curves show the excess of the 1\% top scores, obtained from the bin-per-bin difference between the step and filled orange histograms in
  the left panel. Other curves in the middle and right panels show this excess distribution for other networks.}
\label{fig:seeingdep}
\end{figure*}

To probe sensitivity on the $gri$-band seeing, we used the GAMA09H field which shows the largest differences in seeing distributions
compared to the overall footprint. The difference in seeing FWHM between pairs of bands are plotted in Fig.~\ref{fig:seeingdep} and show a good
match for $g$--$i$, but a major secondary peak in GAMA09H for the $r$--$i$ distribution. This peak corresponds to locations where the $r$- and
$i$-band seeing are anticorrelated with broader light profiles in $r$ band. In these cases, isolated LRGs appear surrounded by bluer halos due
to the PSF mismatch, an artefact that closely mimicks strongly lensed images blended with a central deflector. Unsurprisingly, by plotting the
1\% galaxies in GAMA09H obtaining highest network scores (see Fig.~\ref{fig:seeingdep}), it appears that predictions from the ResNet of
\citetalias{canameras21} heavily depend on the local seeing values. The significant excess of strong-lens candidates identified by this network
for $r$--$i$ FWHM above 0.4\arcsec\ suggests that the underlying model tends to identify color gradients in the galaxy light profiles as lensed
arcs. The baseline ResNet shows a similar distribution as the \citetalias{canameras21} ResNet, with a deficit and excess of high scores for
$r$--$i$ FWHMs below and above 0.4\arcsec, respectively. For the baseline ResNet, the bimodal trend also appears in the difference between $g$
and $i$-band seeing FWHMs. The baseline CNN exhibits a similar bimodality but with lower amplitude in both pairs of bands.

We investigated various methods to reduce this bimodality and obtain invariance to seeing FWHMs. A significant improvement was found by simply
adding the $z$ band as fourth frame to the network inputs, as shown in Fig.~\ref{fig:seeingdep} (right panel) for the baseline CNN architecture. The
median seeing of 0.68\arcsec\ in $z$ band is intermediate between the median values in $i$ band (0.58\arcsec) and $gr$ bands ($\simeq 0.76$\arcsec),
providing a straightforward way to help the networks classify color gradients as observational artefacts. Furthermore, providing the PSF cutouts as
inputs together with the $gri$ science frames also proved successful in removing the bimodality. This technique was previously used in supervised
neural networks, e.g. for estimating galaxy structural parameters \citep{li21b}. In our case, we used the baseline ground-truth data set and we
imported the HSC PDR2 PSF at the position of the central galaxy in each cutout. Zero-padding pixels were added to the 42\,$\times$\,42 pixels PSF
frames to match the dimension of science images. We then used our baseline architectures to train a CNN and a ResNet using a six-dimension input
with the $gri$ coadds and corresponding PSF frames. After appending the PSF frames to our test set we found AUROC, TPR$_{\rm 0}$, and TPR$_{\rm 10}$
of 0.9716, 23.8\%, and 46.6\% for the CNN, comparable to our best networks and plotted in Fig.~\ref{fig:seeingdep}, and 0.9375, 22.8\%, and 32.3\%
for the ResNet, respectively. Other alternatives would be to apply random gaussian blurring as \citet{stein22}, or to provide PSF FWHMs as numerical
inputs before the FC layers, similarly to the combination of morphological parameters and multiband images used by \citet{pearson22} for their
searches for galaxy mergers. Both approaches were discarded since the extra blur affected the recall of small-separation SuGOHI lenses, and we
could not obtain competitive performance for CNNs and ResNets combining image and catalog input.

\begin{figure*}
\centering
\includegraphics[width=.99\textwidth]{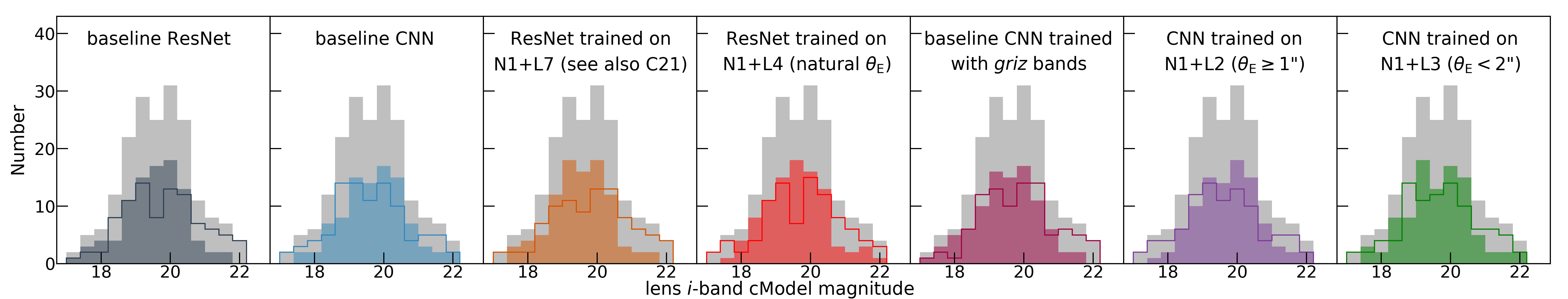}
\includegraphics[width=.99\textwidth]{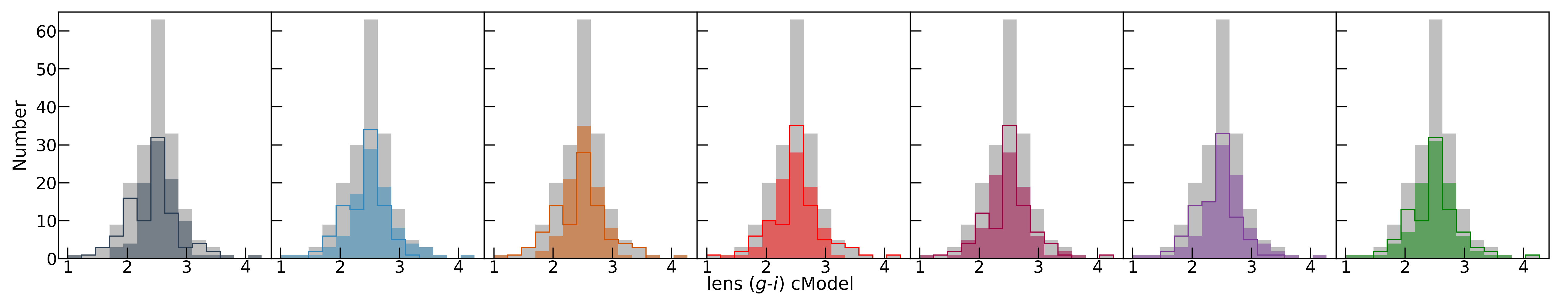}
\includegraphics[width=.99\textwidth]{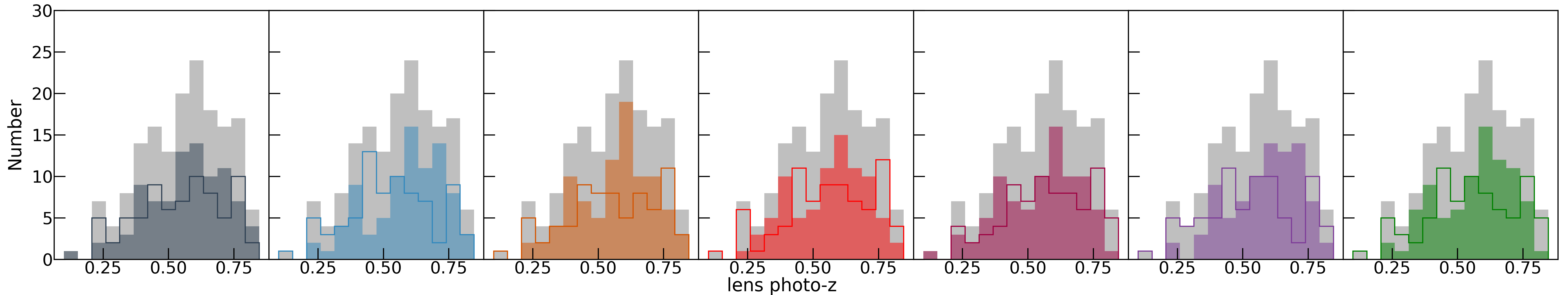}
\includegraphics[width=.99\textwidth]{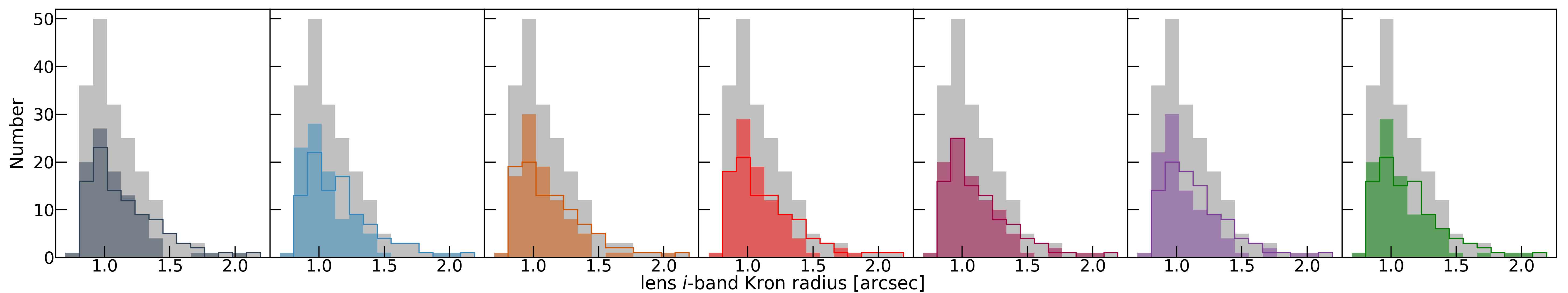}
\includegraphics[width=.99\textwidth]{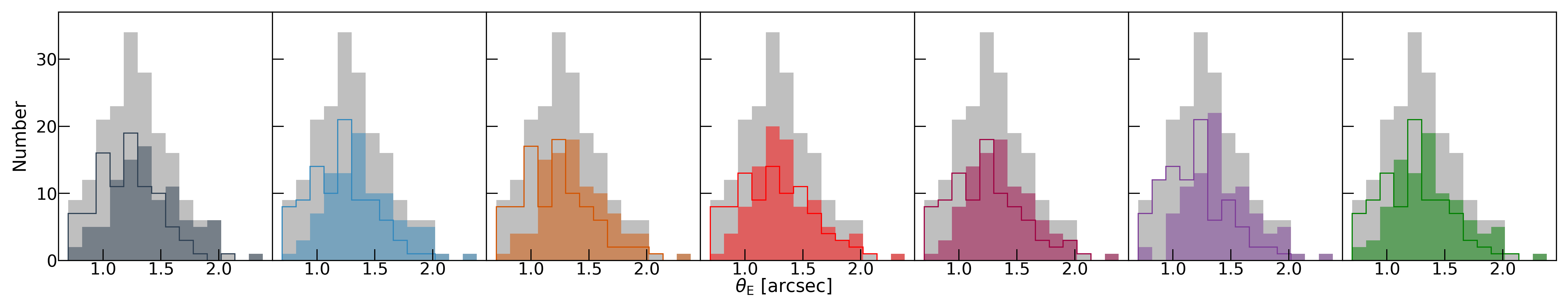}
\caption{Properties of SuGOHI lenses in our test set that are recovered (plain) or missed (steps) by our networks, for fiducial score thresholds
  resulting in 50\% recall. The histograms for all SuGOHI lenses are plotted in light gray. From left to right, we show the baseline ResNet, the
  baseline CNN, the ResNet trained on sets N1$+$L7 (network used in \citetalias{canameras21}), the ResNet trained on sets N1$+$L4 (natural
  $\theta_{\rm E}$ distribution), the baseline CNN trained on $griz$ bands instead of $gri$, the CNN trained on sets N1$+$L2
  ($\theta_{\rm E} \geq 1$\arcsec), and the CNN trained on sets N1$+$L3 ($\theta_{\rm E} < 2$\arcsec).}
\label{fig:sugohiprop}
\end{figure*}

Taking advantage of the on-going status of the HSC Wide survey, we characterized the variations of output scores as a function of image depth.
For each band, the number of frames per stack was used as proxy of depth and we compared the distributions for galaxies over the overall footprint,
within GAMA09H, and with the 1\% highest scores for various networks (see Appendix). While the \citetalias{canameras21} ResNet remains relatively
invariant to the number of frames, the highest scores identified by the baseline CNN shows a strong excess of galaxies with $\leq$3 frames per stack
in all three bands, suggesting that its predictions are biased for the shallowest images. The baseline ResNet shows the opposite trend in $g$ and
$r$ bands, with a deficit and excess of high scores for $\leq$4 and $>$4 frames, respectively. As previously, adding $z$ band as input appears to
naturally avoid biased predictions for galaxies without full-depth observations in $gri$ bands. More homogeneous coverage in later HSC releases and
dedicated observing strategie for {\it Rubin} LSST will also help mitigate such biases in the future.

\subsection{Dependence on image rotation}
\label{ssec:rot}

The variations of network output scores as a function of image orientation were characterized by classifying original HSC cutouts and cutouts
rotated by $k \times \pi/2$, where $k=0$,1,2,3, for the 189 SuGOHI lenses in our test set. The output distibutions of scores show that our neural
networks tend to provide classifications that are nearly-invariant to image orientation for objects identified as clear lenses (mean scores
$\mu_{\rm p} \simeq 1$) and clear nonlenses (mean scores $\mu_{\rm p} \simeq 0$). For instance, we find that predictions with a mean score
$\mu_{\rm p}<0.1$ or $>0.9$ over the four image orientations, systematically have a low average scatter $\sigma_{\rm p} \lesssim 0.02$ for the baseline
CNN, the baseline ResNet, and the ResNet from \citetalias{canameras21}.

SuGOHI lenses with intermediate mean scores are much more sensitive to the orientation of input frames. For lenses with a mean network score
$\mu_{\rm p}$ in the range 0.1--0.9, we find a scatter $\sigma_{\rm p} \simeq 0.18$, 0.21, and 0.16 with the baseline CNN, the baseline ResNet, and
the ResNet from \citetalias{canameras21}, respectively. Using G-CNN architectures allows to significantly lower this dependence. For instance, with
the G-CNN v3 that showed the highest AUROC over the three architectures we tested, the scatter decreases to $\sigma_{\rm p} \simeq 0.08$. Residual
dependencies on image orientation with the group-equivariant architectures are due to the random centering offsets applied to images prior to
classification (see Sect.~\ref{ssec:testaug}).

\subsection{Properties of false negatives and false positives}

As noted in Sect.~\ref{ssec:ens}, there is a substantial overlap between the SuGOHI lenses assigned low scores by different neural networks. In
Fig.~\ref{fig:sugohiprop}, we compare the properties of test lenses that are recovered or missed by a representative selection of seven
high-performing networks. These distributions are plotted for fiducial score thresholds resulting in a common recall of 50\%. For each SuGOHI
system, the cModel magnitudes, the Kron radius, and the photometric redshift \citep[from the Mizuki template-fitting code of][]{tanaka15} of the
foreground lens galaxy are retrieved from the {\tt pdr2\_wide.forced} and {\tt pdr2\_wide.photoz\_mizuki} tables. We also rely on the strong-lens
modeling neural network from \citet{schuldt22a} to obtain the SIE parameters and external shear for all lenses in this test set. The subset of 30
SuGOHI systems with ancillary Markov-Chain Monte-Carlo sampling-based models shows good agreement between both Einstein radius estimates, with the
distribution in $\Delta \theta_{\rm E}$ having median and 1$\sigma$ ranges of $0.06_{-0.06}^{+0.17}$
\citep[see][for a detailed comparison of modeling approaches]{schuldt22b}.

Figure~\ref{fig:sugohiprop} shows that networks miss the majority of lenses with faint deflectors having $i \simeq 21$--22~mag. These networks
are trained on various sets of mocks produced from the same parent sample of LRGs with $z_{\rm spec}$ and $v_{\rm disp}$ measurements. In consequence,
while the $i_{\rm cModel}$ distributions for the mocks and SuGOHI sample both peak at $\simeq 19$--20~mag, all $i>21$~mag deflectors without SDSS
spectroscopy follow-up are discarded from training. The lens ($g-i$) colors do not have such a discrepancy between the training and test lenses.
The recall is roughly constant as a function of lens color, apart from the tail of bluer lens galaxies with ($g-i$) $\lesssim 2$ which have lower
recall in 4/7 networks. In terms of lens $z_{\rm phot}$, we also obtain a stable recall up to the highest lens redshifts of $z_{\rm phot} \gtrsim 0.7$.
This stability was however only reached after boosting the fraction of lens LRGs with $z_{\rm phot}$ above the peak of the SDSS sample in our various
training sets. The baseline CNN is the network showing the most significant drop in recall at $z_{\rm phot} \simeq 0.5$. The distributions of lens
$i$-band Kron radii have stronger differences, especially at the high-end $>1.3$\arcsec, where the majority of lenses are missed by all seven
networks. In contrast to $i$-band magnitudes, this bias results from the acceptance criteria of simulated arcs rather than properties of the parent
LRG sample. Besides excluding mocks with multiple images buried in lens light, the $R_{\rm sr/ls,min}$ threshold also tends to discards extended LRGs
which dominate over lensed source emission for any value of $\mu$. This explains the failure in identifying this subset of SuGOHI lenses.

The recall of SuGOHI lenses strongly varies with Einstein radius. Even though simulations for training the baseline networks and the ResNet of
\citetalias{canameras21} cover the entire 0.75--2.5\arcsec\ range uniformly, the recall at $\theta_{\rm E} \lesssim 1.2$\arcsec\ is $<$15\%. The
recovery of wider image separation systems is higher, but still incomplete. Fig.~\ref{fig:sugohiprop} also includes networks trained on natural
$\theta_{\rm E}$ distribution, and on uniform distributions with $\theta_{\rm E} \geq 1$\arcsec\ and $\theta_{\rm E} < 2$\arcsec. These networks do
not succeed in extracting blended lensed arc features, but we nonetheless notice an influence from the relative fraction of compact and extended
image configurations in the training sets, as the recall at $\theta_{\rm E} \lesssim 1.2$\arcsec\ increases for the ResNet trained mostly on 
low-$\theta_{\rm E}$ mocks. Interestingly, the recall at small $\theta_{\rm E}$ does not improve significantly for the network trained on difference
images (not shown in Fig.~\ref{fig:sugohiprop}).

We also classified the SuGOHI colors visually to assign a ``red'' flag to sources redder than the lens galaxy. We found that test sample is
strongly dominated by blue lensed sources, with only 10 sources marked as ``red''. Despite our efforts in extending the source color distribution
in the training sets, we find that the recall of red SuGOHI arcs remains low. Only the baseline ResNet, the ResNet trained on sets N1$+$L4, and
the CNN trained on four bands are able to find three or four of these 10 sources, while other networks miss them all. This matches results for
other lens-finders \citep[e.g.,][]{jacobs22}.

Overall, misclassified lenses tend to have properties deviating from the bulk of the mocks in the training sets (see e.g., Fig.~\ref{fig:fns_resnet}
for the baseline ResNet). This common behaviour for supervised machine learning algorithms can be corrected by expanding the parameter space
covered by the simulations. In the future, deeper spectroscopic surveys will help in that regard, as they will provide more diverse lens and source
samples for our simulation pipeline. Regarding false positives, we found that high-performing networks are contaminated by diverse, mostly distinct
galaxy types. Only a few of them show apparent arc-like features, and the majority belong to morphological classes relatively less represented in
our ground-truth data, such as galaxies with spiral lanes, or edge-on disks. Finally, contaminants do not follow obvious trends with structural 
parameters.

\section{Summary}
\label{sec:conclu}

We conducted a systematic comparison of supervised neural networks for selecting galaxy-scale strong gravitational lenses in ground-based imaging.
Identifying the main ingredients to optimize the classification performance is becoming crucial to reduce contamination rates and the need in human
resources. We used PDR2 images from the HSC Wide survey to address this issue and to prepare for the exploitation of forthcoming deep, wide-scale
surveys such as {\it Rubin} LSST. A representative test set was designed with 189 strong lenses previously found in HSC, and 70,910 nonlens galaxies
in the COSMOS field including realistic number and diversity of lens-like galaxies to mimick an actual classification set up. Multiple networks were
trained on different sets of realistic strong lens simulations and nonlens galaxies, with various architectures and data pre-processing, and mainly
using the $gri$ bands that have optimal depth. These networks reached excellent AUROC on the test set, but the recall for zero and 10 false positives
showed wide ranges of values. The following conclusions are drawn:
\begin{itemize}
\item As expected, choices in the construction of the ground-truth training data have a major impact. Optimal performance are only obtained for data
  sets comprising mock lenses with bright, deblended multiple images, and including boosted fractions of nonlens contaminants. The AUROC, TPR$_{\rm 0}$,
  and TPR$_{\rm 10}$ are typically, but not systematically, higher for the baseline ResNet. The metrics are less affected by changes in the ground-truth
  data for the baseline CNN.
\item The best performance are obtained for the baseline ResNet which is adapted from ResNet18. We explored variations around the baseline CNN, ResNet,
  and G-CNN architectures finding that, while some CNNs perform better than ResNets, TPR$_{\rm 0}$ tend to be higher for ResNets ($\simeq$ 10--40\%), with
  none of the CNNs exceeding TPR$_{\rm 0}$ $\simeq$ 10\%.
\item Among the data pre-processing that we tested, applying random shifts to the image centroids, and square root stretches to the pixel values appear
  to be most helpful. Simple augmentation procedures, such as loading the frames mirrored horizontally and vertically together with the original images,
  help increase the recall at low contamination for the CNN, but not the ResNet. For CNNs, using random viewpoints of the original images as input also
  provide an additional gain compared to using 10\arcsec\,$\times$\,10\arcsec\ $gri$ stacks. We also find an improvement when adding $z$ band together
  with $gri$.
\item The CNN shows more stable performance than the ResNet as a function of training set size, when varying between 7000 and 70,000 examples. The
  metrics, TPR$_{\rm 0}$ in particular, raise significantly for the ResNet when increasing the number of training examples.
\item Using $g - \alpha i$ difference images to subtract emission from the main central galaxy does not improve the classification, likely due to the
  differences in $g$ and $i$-band seeing. However, we find that masking neighboring galaxies leads to a major gain in AUROC for the CNN.
\item Using committees of networks trained on different data sets, and with the lowest overlap in false positives, turns out to provide the most
  significant gain in performance. This results in the highest TPR$_{\rm 0}$ and TPR$_{\rm 10}$ over our various tests of about 60\% and 80\%, respectively.
  In addition, using committees of networks with different weight initialisation does not provide such a gain, but helps stabilize the metrics compared
  to individual networks.
\end{itemize}
Moreover, despite using accurate PSF models in the lens simulations, some networks show systematic dependence with variations in seeing FWHM and image
depth between observing bands over the footprint. For our baseline networks trained on $gri$ bands, dependencies on image quality are particularly
important over regions where the $r$- and $i$-band seeing are anticorrelated, with broader light profiles in $r$ band. The underlying models tend to
associate such color gradients produced by the PSF mismatch as genuine strong lensing features, leading to an excess of spurious strong-lens candidates.
We investigated various methods to obtain invariance to seeing FWHMs, and found excellent improvement when training either with the four $griz$ bands,
or with $gri$ science frames together with the corresponding PSF cutouts. This illustrates that specific strategies are needed to ensure neural
networks correctly account for observational effects in their underlying models.

Overall, the systematic tests presented in this paper demonstrate the feasibility to reach a recall at zero contamination as high as 60\%, which opens
promising perspectives for pure selection of strong lenses without or with modest human input. While the conclusions drawn from this analysis are
necessarily affected by the construction of our test sets, and by potential fluctuations in TPR$_{\rm 0}$ given the finite sample size, we expect they
will hold for next generation ground-based surveys with comparable image depth and quality, such as {\it Rubin} LSST. Together with new strong-lens
simulations matching the image quality of these forthcoming surveys, and with extensive tests of the network selection functions (e.g., More et al.,
in prep.), these ingredients will allow optimal classification performance to be reached.

In the future, one priority will be to increase the recall of strong lenses with broader types of lens potentials, including exotic systems that are
more difficult to mock up \citep[see, e.g.,][]{wilde22}, while maintaining low contamination rates. This could be done via novel approaches, such as
combinations of unsupervised and supervised machine learning methods, or anomaly detection algorithms. \citet{storey21} explored anomaly detection
in the same HSC data set, by training a generative adversarial network on HSC Wide PDR2 multiband images. While 4500 out of their 13000 high-anomaly
objects are included in our parent sample of galaxies with Kron radius $\geq 0.8$\arcsec, only a handful are flagged as strong-lens candidates in
\citetalias{canameras21}, likely due to diffraction spikes from nearby stars rather than the presence of lensed arcs. Further work is therefore needed
to identify the rare strong gravitational lenses as systems deviating significantly from standard morphological classes.

\section*{Acknowledgements}

R.C. and S.H.S. thank the Max Planck Society for support through the Max Planck Research Group for S.H.S. This project has received funding from the
European Research Council (ERC) under the European Union's Horizon 2020 research and innovation program (LENSNOVA: grant agreement No 771776). This
research is supported in part by the Excellence Cluster ORIGINS which is funded by the Deutsche Forschungsgemeinschaft (DFG, German Research Foundation)
under Germany's Excellence Strategy -- EXC-2094 -- 390783311. S.S. acknowledges financial support through grants PRIN-MIUR 2017WSCC32 and 2020SKSTHZ.
S.B. acknowledges the funding provided by the Alexander von Humboldt Foundation. A.T.J. is supported by the Program Riset Unggulan Pusat
dan Pusat Penelitian (RU3P) of LPIT Insitut Teknologi Bandung 2023. This paper is based on data collected at the Subaru Telescope and retrieved from
the HSC data archive system, which is operated by Subaru Telescope and Astronomy Data Center at National Astronomical Observatory of Japan. The Hyper
Suprime-Cam (HSC) collaboration includes the astronomical communities of Japan and Taiwan, and Princeton University. The HSC instrumentation and
software were developed by the National Astronomical Observatory of Japan (NAOJ), the Kavli Institute for the Physics and Mathematics of the Universe
(Kavli IPMU), the University of Tokyo, the High Energy Accelerator Research Organization (KEK), the Academia Sinica Institute for Astronomy and
Astrophysics in Taiwan (ASIAA), and Princeton University. Funding was contributed by the FIRST program from Japanese Cabinet Office, the Ministry of
Education, Culture, Sports, Science and Technology (MEXT), the Japan Society for the Promotion of Science (JSPS), Japan Science and Technology Agency
(JST), the Toray Science Foundation, NAOJ, Kavli IPMU, KEK, ASIAA, and Princeton University. This paper makes use of software developed for the LSST.
We thank the LSST Project for making their code available as free software at http://dm.lsst.org. This work uses the following software packages:
{\tt Astropy} \citep{astropy13,astropy18}, {\tt matplotlib} \citep{hunter07}, {\tt NumPy} \citep{vanderwalt11,harris20}, {\tt Scipy} \citep{virtanen20},
{\tt torch} \citep{paszke19}.

\bibliography{lenssearch}

\begin{appendix}

\section{Receiver operating characteristic curves}
  
\begin{figure}
\centering
\includegraphics[width=.49\textwidth]{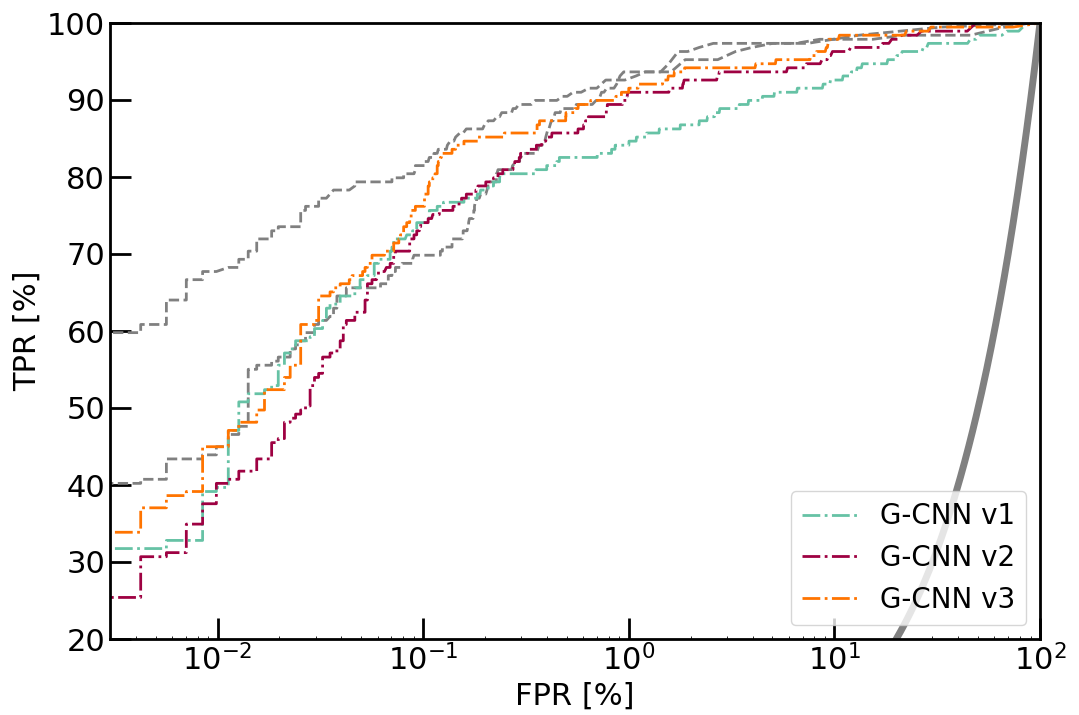}
\caption{Influence of the network architecture, for our baseline data set and the group-equivariant network architectures G-CNNs adapted from
  \citet{cohen16}. For reference, the dashed grey lines show two good networks (the baseline ResNet and the ResNet from C21). The thick grey curve
  corresponds to a random classifier.}
\label{fig:archi}
\end{figure}

\begin{figure}
\centering
\includegraphics[width=.49\textwidth]{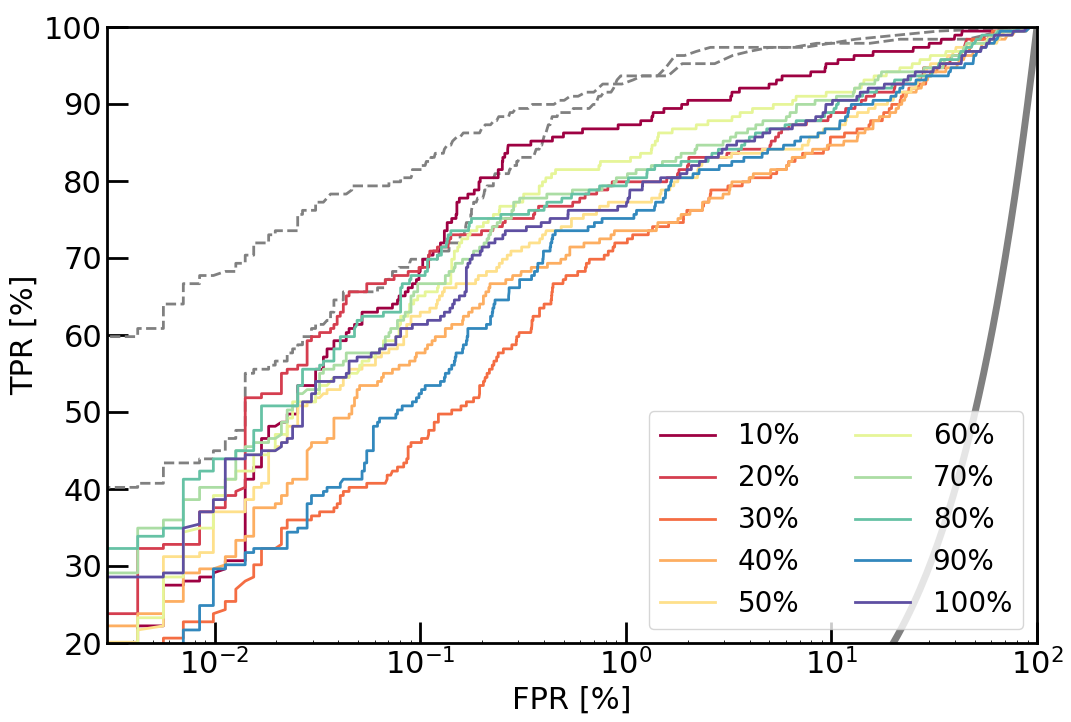}
\includegraphics[width=.49\textwidth]{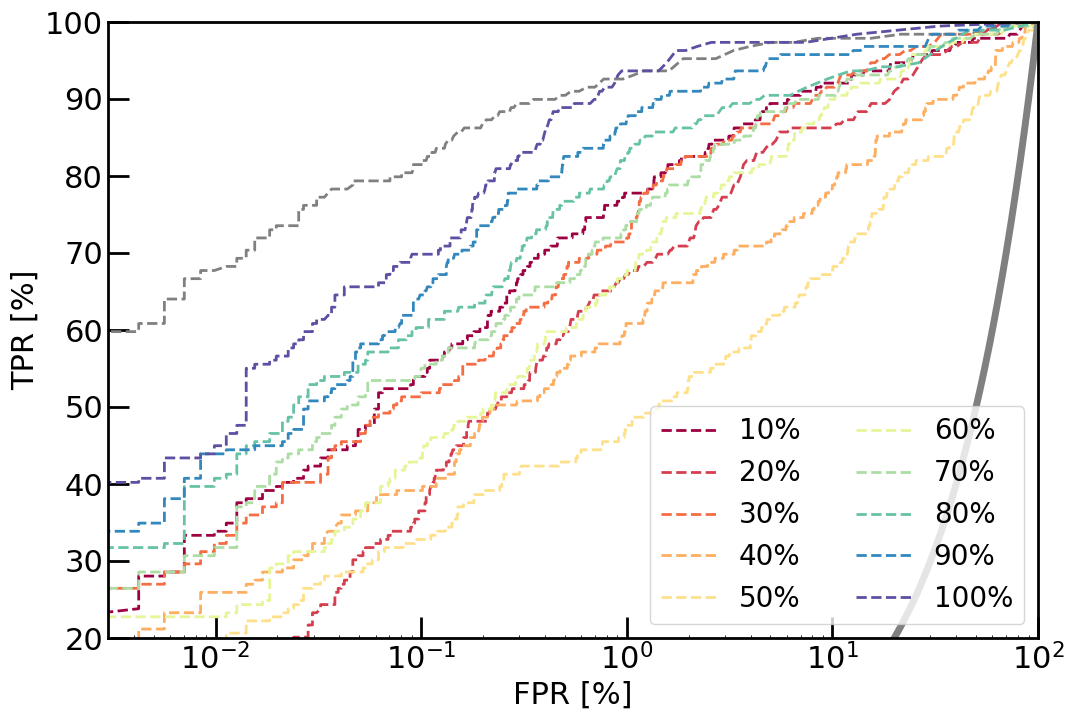}
\caption{ROC curves for training with different fractions of the overall data set, for the baseline training set and for the baseline CNN (top) and
  ResNet (bottom). For reference, the dashed grey lines show two of the best networks from Fig.~\ref{fig:dataset} (the baseline ResNet and the ResNet
  from C21). The thick grey curve corresponds to a random classifier.}
\label{fig:datasize}
\end{figure}

\section{Dependence on image depth}

\begin{figure*}
\centering
\includegraphics[height=.24\textwidth]{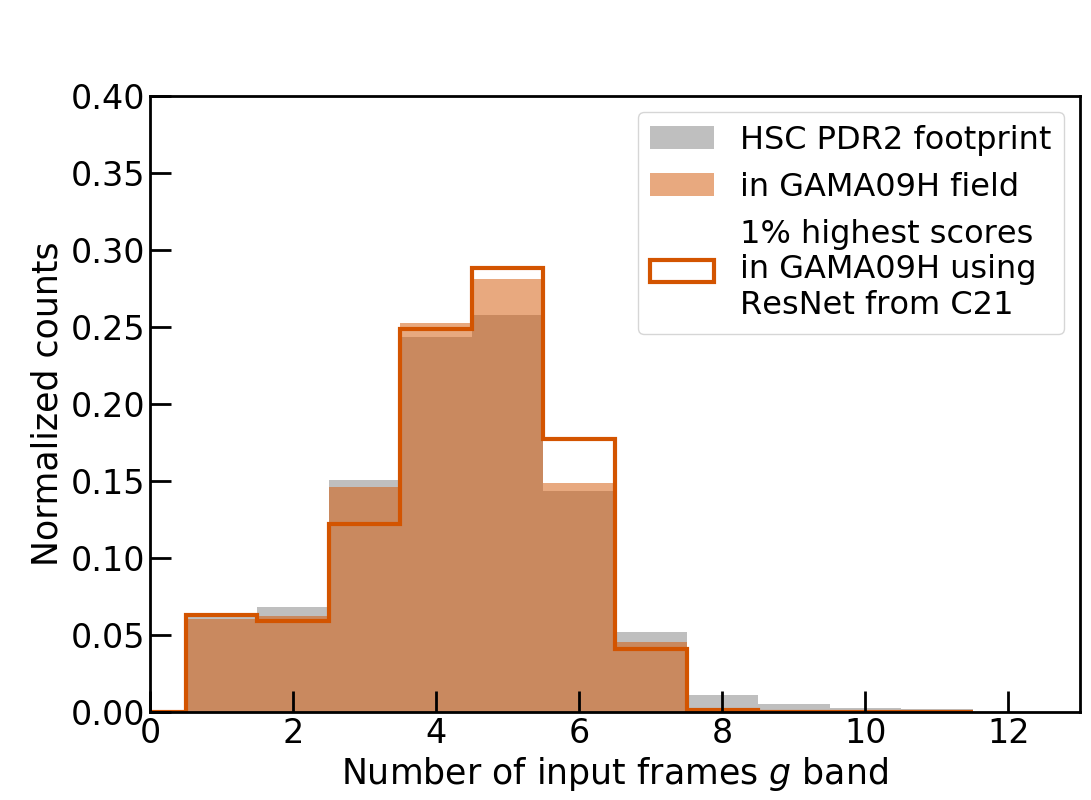}
\includegraphics[height=.24\textwidth]{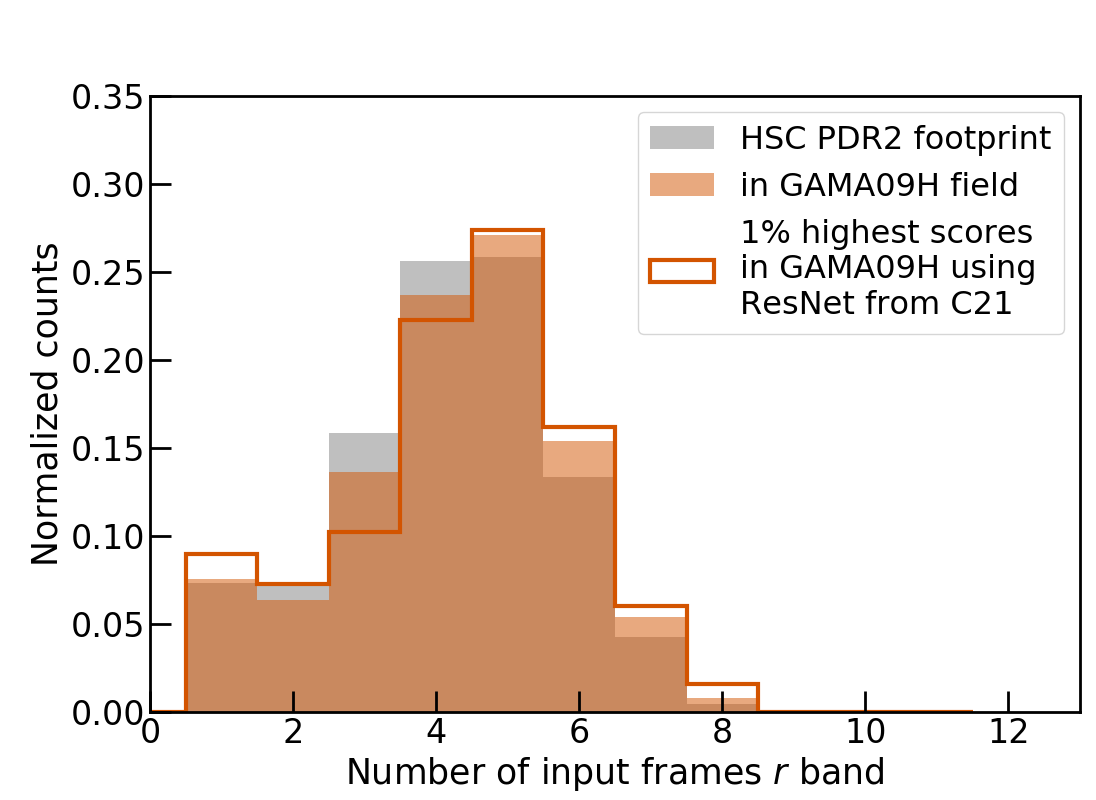}
\includegraphics[height=.24\textwidth]{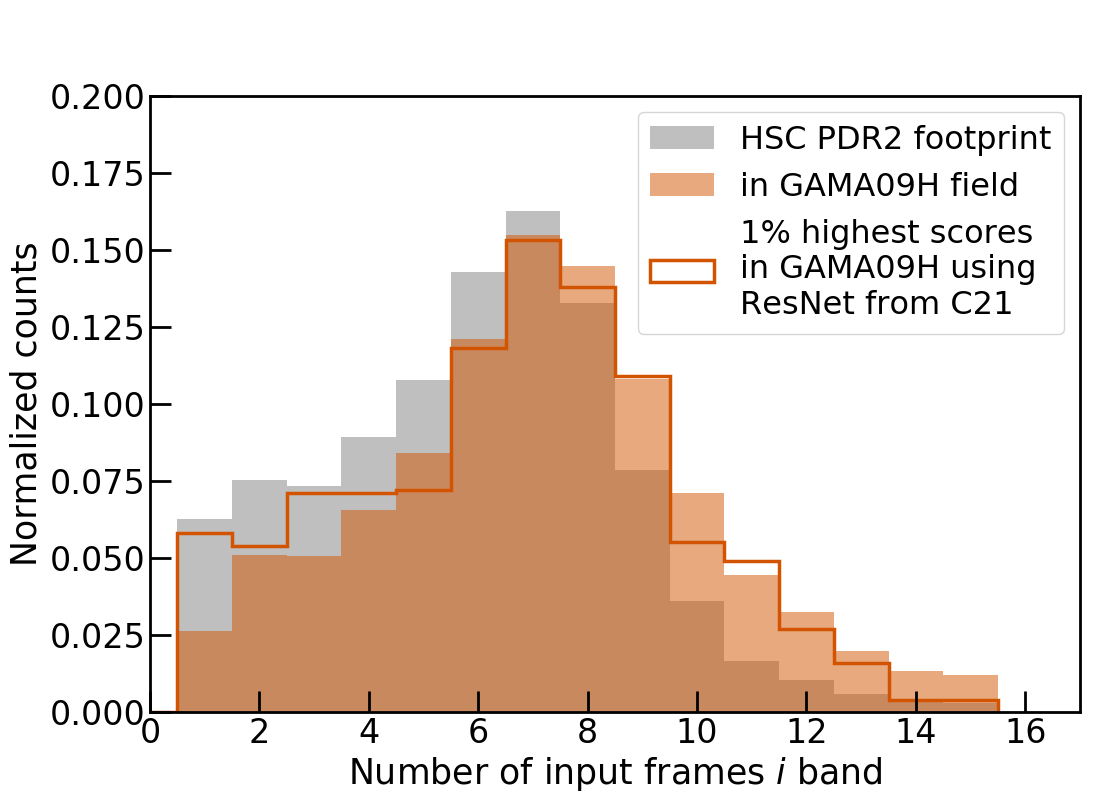}

\includegraphics[height=.24\textwidth]{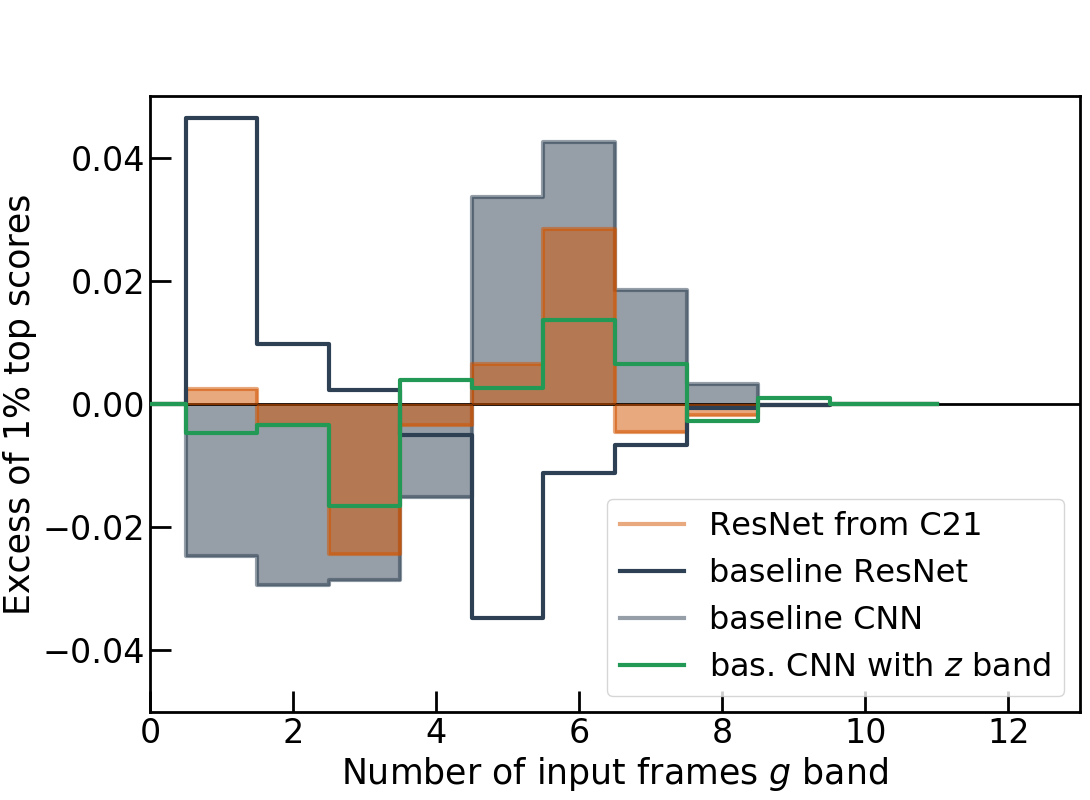}
\includegraphics[height=.24\textwidth]{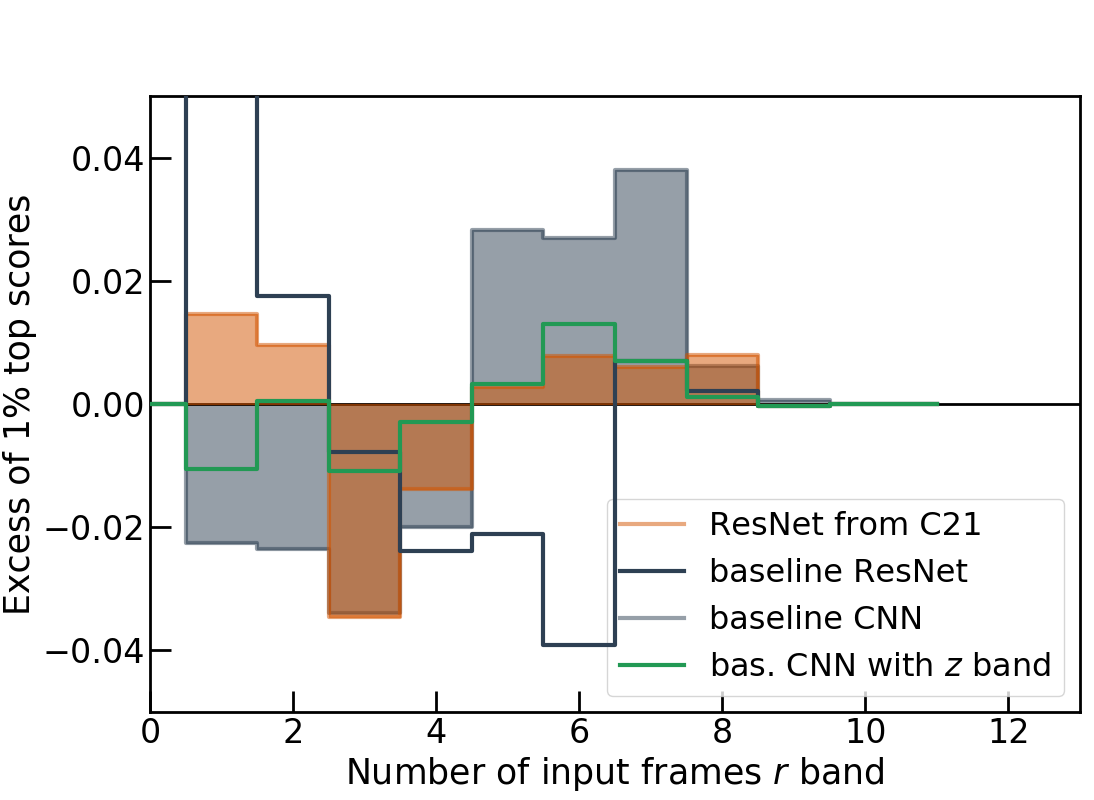}
\includegraphics[height=.24\textwidth]{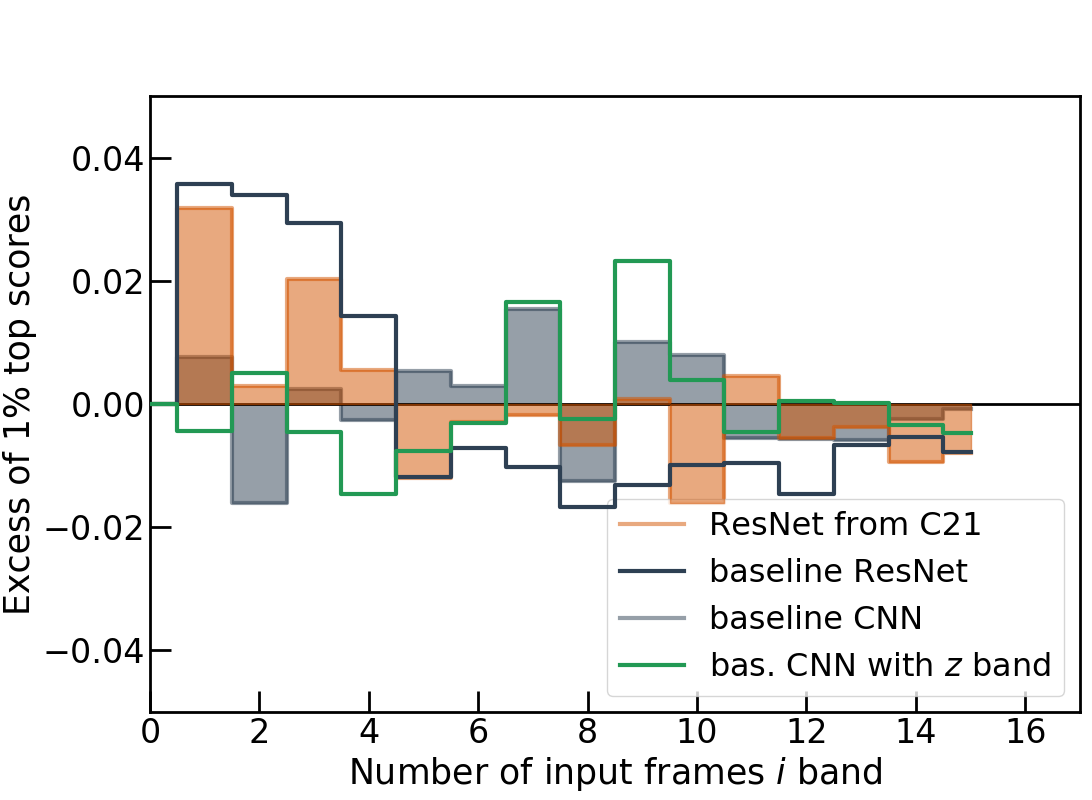}
\caption{Histograms of the number of frames per stack in each band. In the top panels, grey histograms show the average distributions over the
  entire HSC PDR2 footprint, and orange filled histograms show the distributions restricted to the GAMA09H field. In addition, step histograms show
  the distributions for galaxies within GAMA09H assigned the 1\% highest scores by the ResNet from \citetalias{canameras21}. In the bottom panels,
  orange curves show the excess of the 1\% top scores, obtained from the difference between the step and filled orange histograms in the top panels,
  and other curves showing additional networks.}
\label{fig:depthdep}
\end{figure*}

\begin{figure*}
\centering
\includegraphics[width=.48\textwidth]{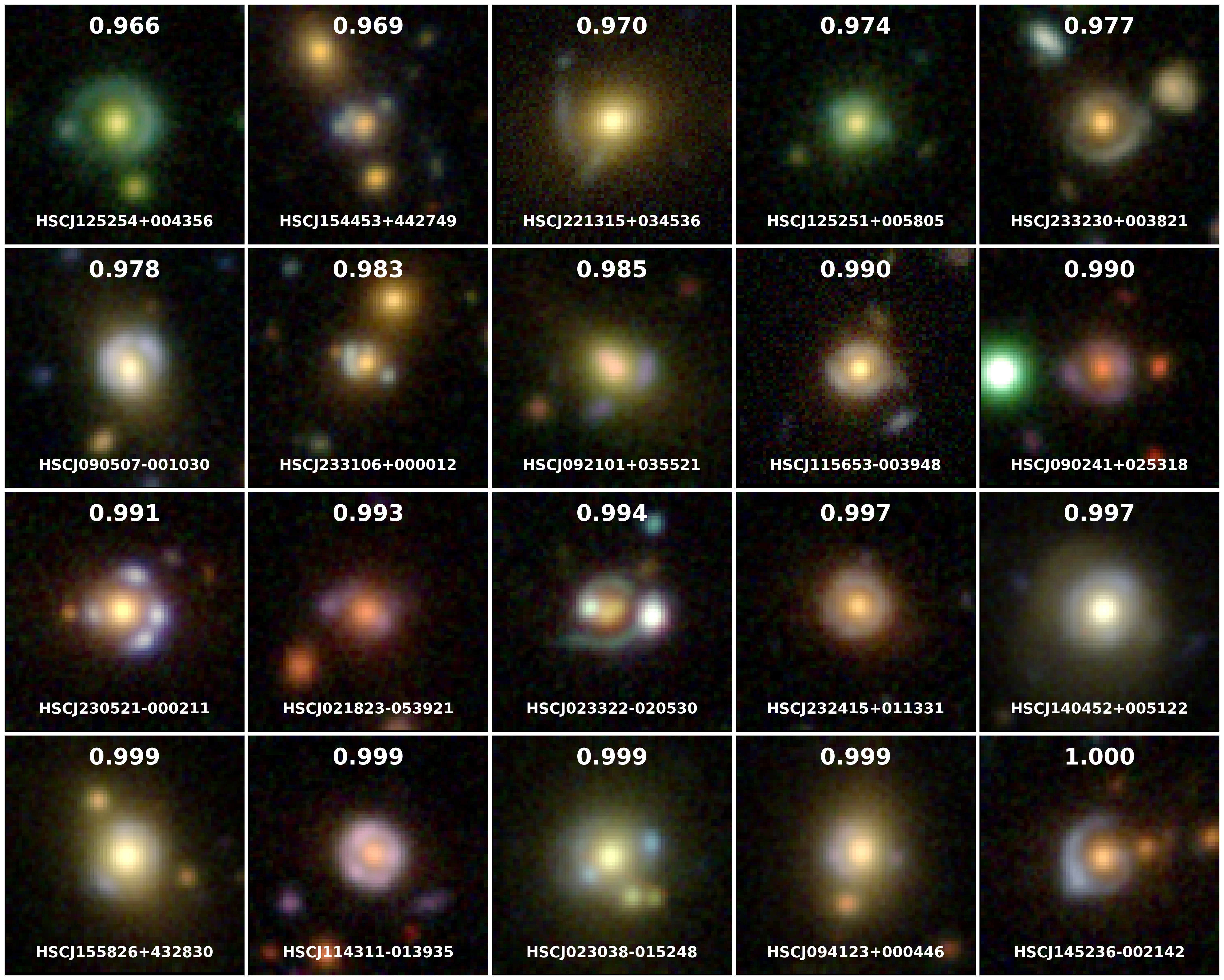}
\hspace{3mm}
\includegraphics[width=.48\textwidth]{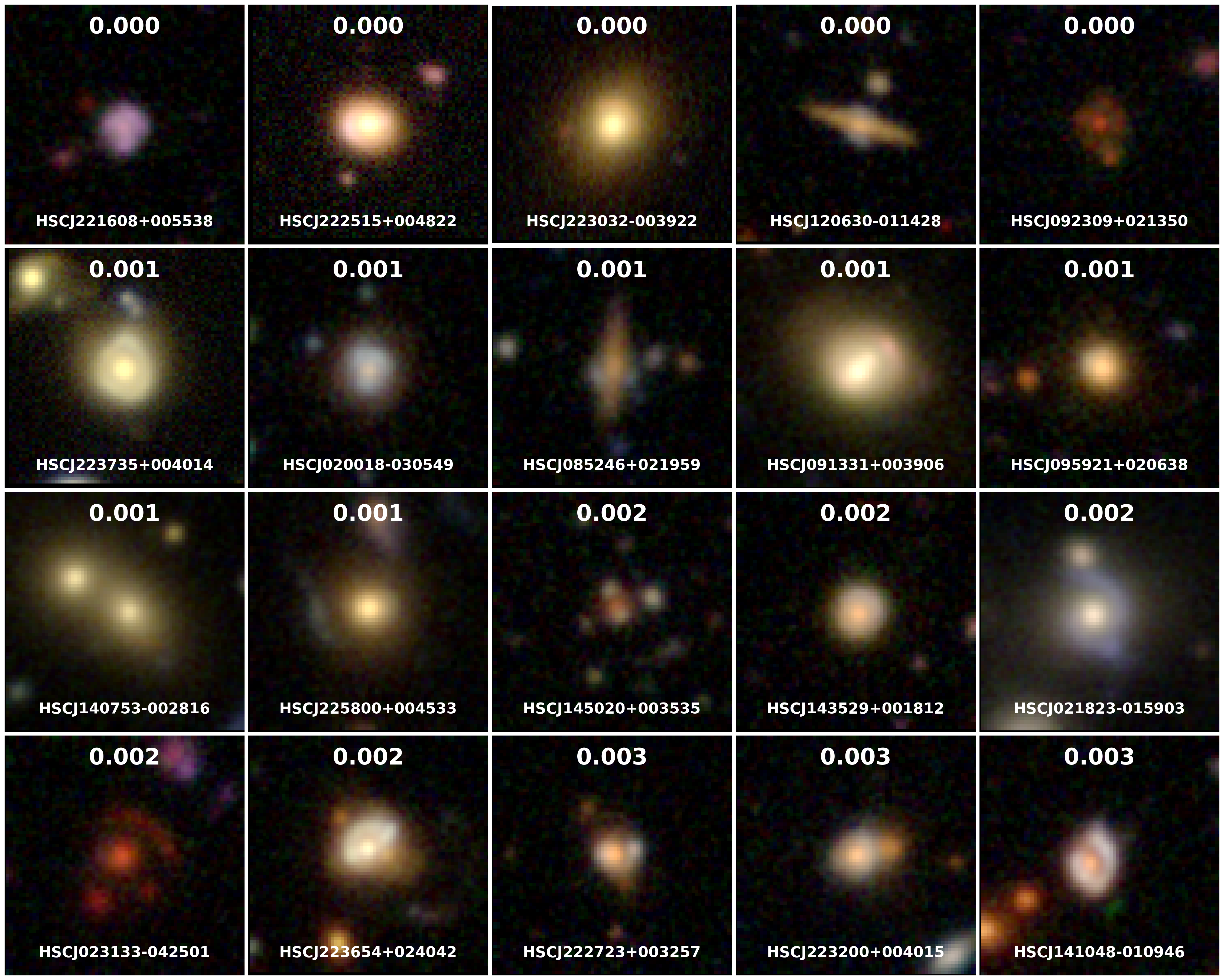}
\caption{Examples of 20 SuGOHI lenses recovered (left) and missed (right) by our baseline ResNet ranked by increasing network scores.}
\label{fig:fns_resnet}
\end{figure*}

\end{appendix}

\end{document}